# The influence of Structural Dynamics in Two-Dimensional Hybrid Organic-Inorganic Perovskites on their Photoluminescence Efficiency – Neutron scattering analysis

Haritha Sindhu Rajeev, [1,*] Xiao Hu, [1,2] Wei-Liang Chen, [3] Depei Zhang, [1,4] Tianran Chen, [1,5] Maiko Kofu, [6] Ryoichi Kajimoto, [6] Mitsutaka Nakamura, [6] Alexander Z. Chen, [7] Grayson C. Johnson, [7] Mina Yoon, [8] Yu-Ming Chang, [3] Diane A. Dickie, [9] Joshua J. Choi, [7] and Seung-Hun Lee [1]

[1] *Department of Physics, University of Virginia, Charlottesville, Virginia 22904, USA*

[2] *Condensed Matter Physics and Materials Science Division,*

*Brookhaven National Laboratory, Upton, NY 11973, USA*

[3] *Center for Condensed Matter Sciences, National Taiwan University, Taipei 10617, Taiwan*

[4] *Currently working in Robinhood, USA*

[5] *Department of Physics and Astronomy, University of Tennessee, Knoxville TN 37996, USA*

[6] *J-PARC Center, Japan Atomic Energy Agency, Ibaraki 319-1195, Japan*

[7] *Department of Chemical Engineering, University of Virginia, Charlottesville, Virginia 22903, USA*

[8] *Center for Nanophase Materials Sciences, Oak Ridge National Laboratory, Oak Ridge TN 37831, USA*

[9] *Department of Chemistry, University of Virginia, Charlottesville, Virginia 22904, USA*

**Abstract**

Two-dimensional hybrid organic-inorganic perovskites (HOIPs) have emerged as promising materials for light-emitting diode applications. In this study, by using time-of-flight neutron spectroscopy we identified and quantitatively separated the lattice vibrational and molecular rotational dynamics of two perovskites, butylammonium lead iodide $(BA)_2PbI_4$ and phenethyl-ammonium lead iodide $(PEA)_2PbI_4$. By examining the corresponding temperature dependence, we found that the lattice vibrations, as evidenced by neutron spectra, are consistent with the lattice dynamics obtained from Raman scattering. We revealed that the rotational dynamics of organic molecules in these materials tend to suppress their photoluminescence quantum yield (PLQY) while the vibrational dynamics did not show predominant correlations with the same. Additionally, we observed photoluminescence emission peak splitting for both systems, which becomes prominent above certain critical temperatures where the suppression of PLQY begins. This study suggests that the rotational motions of polarized molecules may lead to a reduction in exciton binding energy or the breaking of degeneracy in exciton binding energy levels, enhancing non-radiative recombination rates, and consequently reducing photoluminescence yield. These findings offer a deeper understanding of fundamental interactions in 2D HOIPs and could guide the design of more efficient light-emitting materials for advanced technological applications.

## 1. INTRODUCTION

Metal halide perovskites (MHPs) have drawn tremendous research interest due to their wide range of applications, such as next-generation solar cells [1-6], light-emitting diodes (LEDs) [7-9], lasers [10, 11], and photodetectors [12, 13]. These materials are characterized by extended carrier lifetimes, long carrier diffusion lengths, and exceptional carrier protection from defects. Within the MHP family, hybrid organic-inorganic perovskites (HOIPs) are particularly notable for their photovoltaic and optoelectronic properties, attributed to the presence of polarized organic molecules. For the three-dimensional (3D) HOIPs, experimental evidence suggests that the reorientation of the polarized molecules aids polaron formation [14-17]. These polarons, which are quasiparticles formed by excess charge carriers (electrons or holes) and the polarized molecules via Coulomb interaction [18], enhance the screening effects on charge carriers and hence prolong the charge carrier lifetime [19]. Conversely, pure inorganic MHPs, despite lacking organic molecules, have also demonstrated moderate photovoltaic performance [20, 21], highlighting the significant role of the inorganic perovskite framework vibrations through their interactions with the charge carriers. In general, there are two scenarios for phonon-mediated polaron formation: (1) polaron formation facilitated by optical phonons [14, 22]. Free charge carriers interact with polarized optical phonons via Coulomb interaction and get screened or protected from defects and impurities. (2) overdamping of acoustic phonons [23-25]. Most of the excess energy of the photo-excited charge carriers gets dissipated via scattering from acoustic phonons. Upon heating, acoustic phonons get overdamped, which reduces their interactions with charge carriers. Both scenarios allow charge carriers to survive for a longer time.

On the other hand, in two-dimensional (2D) HOIPs, the polarized organic molecules, together with the inorganic framework, establish a layered quantum well structure [26-30], where the organic molecule layer serves as a potential 'wall,' while the inorganic layer serves as a potential 'well.' Instead of polarons in 3D HOIPs, charge carriers confined in this 2D HOIP quantum well tend to form excitons, which are quasiparticles made up of opposite charge carriers with large binding energies (a few hundred meV) [31]. Upon photoexcitation, large numbers of stable excitons accumulate around recombination centers, such as charge point defects, which significantly increase the radiative recombination rate and lead to enhanced photoluminescence quantum yield (PLQY) [26, 30].

However, the emergent microscopic mechanisms that play a significant role in the photoluminescence performance of these 2D HOIPs have not been comprehensively studied. The impacts of inorganic layers are often emphasized, while the roles of polarized molecules appear to be underexplored and underestimated. Previous studies [32] on these soft lattice perovskites showed that the valence band maximum (VBM) is determined by the halide $p$-orbital hybridized with metal $s$-orbital and the conduction band minimum (CBM) is mainly contributed from metal $p$-orbitals, which reflects the fact that the inorganic framework plays a leading role in the electronic band structure construction and hence optoelectronic properties. For example, Gong and his colleagues [33] reported that the PLQY of 2D HOIPs could be influenced by electron-phonon interactions with different crystal rigidity levels. They characterized the role of crystal rigidity in two 2D HOIPs, $(BA)_2PbBr_4$ and $(PEA)_2PbBr_4$ by comparing atomic displacements, spin-lattice relaxation, and time-variation in the electronic band structure. They proposed lower crystal rigidity and a much stronger electron-phonon interaction in $(BA)_2PbBr_4$ to account for its pronounced drop of PLQY at room temperature. Nevertheless, the general crystal rigidity cannot well explain their temperature-dependent measurements of PLQY (Fig. S9(a) in Ref. [33]) -- Below 150 K, both materials persist a high-level PLQY ( $> 90\%$ ); upon further heating up to room temperature, the PLQY of $(BA)_2PbBr_4$ dramatically drops down to $17\%$ while that of $(PEA)_2PbBr_4$ remains above $70\%$. The intriguing behavior of PLQY indicates the activation of some hidden mechanism around 150 K in $(BA)_2PbBr_4$, which our work proposed to be a rotational motion of the polarized organic molecule. In any case, a detailed comparative study among lattice dynamics, molecular rotational dynamics, and photoluminescence performance is essential to complete the big picture of 2D HOIP optoelectronics.

In this work, we focus on two iodine counterparts, $(BA)_2PbI_4$ and $(PEA)_2PbI_4$, which would have similar optoelectronic properties with the corresponding bromides. We have used temperature-dependent X-ray diffraction to characterize their crystal structures and octahedral distortions, which could be closely related to variations in electronic band structure. We used Quasi-Elastic Neutron Scattering (QENS) to directly probe the rotational motion of hydrogen-contained molecules as a function of temperature, from which we have quantitatively identified the rotational modes of different rotors in these two materials with the help of group theory and jump model analysis [34, 35]. Moreover, temperature-dependent Inelastic Neutron Scattering (INS) and Raman scattering measurements were conducted to systematically analyze relevant lattice vibrations (i.e., phonons). All these results were compared with the temperature-dependent evolution of photoluminescence performance. Based on our comparison, we suggest that in these two 2D HOIPs, increasing temperature activates and enhances the rotational motion of polarized organic molecules, which inversely correlates with the temperature-dependent variation of PLQY, while lattice vibrations do not exhibit a clear relationship. We thus suggest a scenario that the rotational motion of polarized organic molecules works as dynamical perturbations to the inorganic framework which fundamentally dominates the construction of electronic band structure. These perturbations interfere with the dielectric environment surrounding confined excitons, potentially reducing the exciton binding energy or breaking binding energy degeneracy, enhancing the non-radiative decay of charge carriers, and thereby suppressing the PLQY.

## 2. EXPERIMENTAL DETAILS

Powder X-ray diffraction (XRD) from 80 K to 300 K was at the University of Virginia on a Bruker D8 VENTURE dual-wavelength Mo/Cu Kappa four-circle diffractometer equipped with an Oxford Cryostream 800 low-temperature device. The 2-D XRD measurements were collected as a series of correlated phi scans which were integrated using the XRD[2]Eval plugin within the Bruker APEX5 software suite [36]. Quasi-elastic neutron scattering (QENS) measurements were conducted on the cold neutron disk chopper spectrometer AMATERAS [37] at the Japan Proton Accelerator Research Complex (J-PARC) with incident energies $E_i = 3.3$meV and 8 meV with an energy resolution of $\sim 1\%$ of $E_i$. Inelastic neutron scattering data were collected on the 4D Space Access Neutron Spectrometer (4SEASONS) [38, 39], at J-PARC. A series of incident neutron energies ($E_i = 10, 30, 62, 115, 280$, and $712$ meV) were selected to cover the entire phonon spectra upon heating from 8 K to 300 K with an energy resolution of $\sim 5\%$ of $E_i$. Raman scattering and photoluminescence measurements were performed at the Center for Condensed Matter Sciences, National Taiwan University. To obtain further insights into the lattice vibrations, density-functional theory (DFT) calculations were performed using the Vienna Ab-initio simulation package (VASP) [40] with the projected augmented wave method [41] and Perdew-Burke-Ernzerhof exchange-correlation potential [42]. For the plane wave basis set, 400 eV cut-off energy was used.

## 3. RESULTS AND ANALYSIS

### 3.1. X-ray diffraction

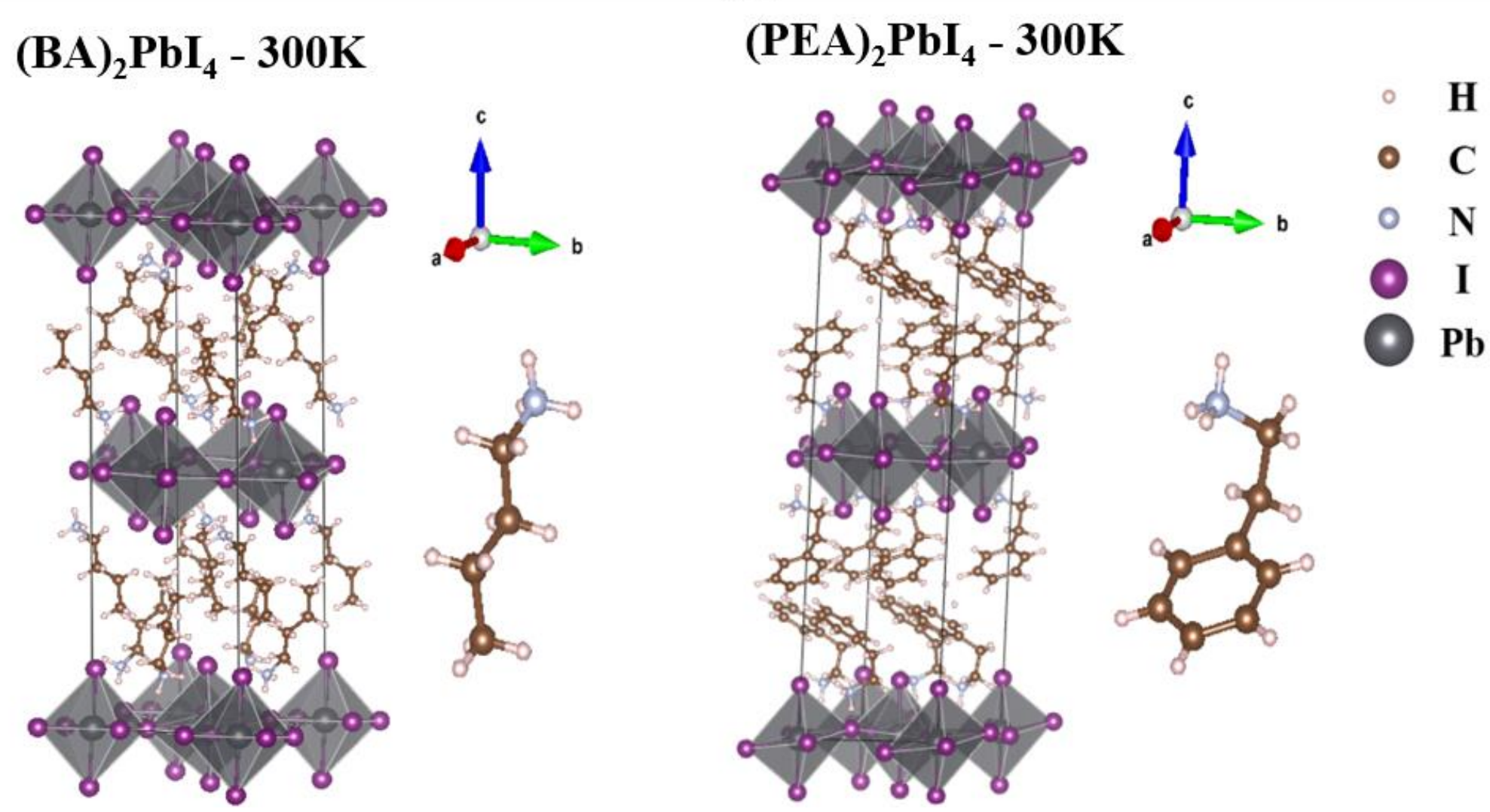

**FIG. 1** Crystal structure of $(BA)_2PbI_4$ and $(PEA)_2PbI_4$ generated using VESTA [43] at 300K based on structural parameters obtained from XRD pattern refinement.

The structures of 1-layer lead iodide perovskite $(BA)_2PbI_4$ and $(PEA)_2PbI_4$ are shown in Figure 1. These structures were obtained by refining the X-ray Diffraction (XRD) patterns at 300 K (Fig. 2). In both structures, the inorganic Pb-I layers are intercalated with bulky organic molecules. In $(BA)_2PbI_4$ the BA molecule is a long alkyl chain $CH_3CH_2CH_2CH_2NH_3$, whereas for $(PEA)_2PbI_4$ the PEA molecule has a toluene ($C_6H_5CH_2$) connected to a methylamino group ( $CH_2NH_3$). At room temperature $(BA)_2PbI_4$ crystallizes in the orthorhombic $Pbca$ space group, with lattice constants $a < b$. The system undergoes a structural phase transition at $\sim 275$ K ($T_c$), after which it still crystallizes in the orthorhombic Pbca space group but with lattice constants $a > b$. In contrast, $(PEA)_2PbI_4$ does not undergo a structural transition from 80 K to 300 K and crystallizes in the triclinic $P\bar{1}$ space group. These findings are consistent with other literature studies on these materials [44- 49]. The details of unit cell parameters obtained from the refinement of XRD patterns are shown in Table 1. The initial reference crystal structure for refinement of our XRD patterns were taken from [44] for $(BA)_2PbI_4$ and [47] for $(PEA)_2PbI_4$. See supplementary file figure S1 for more refinement details.

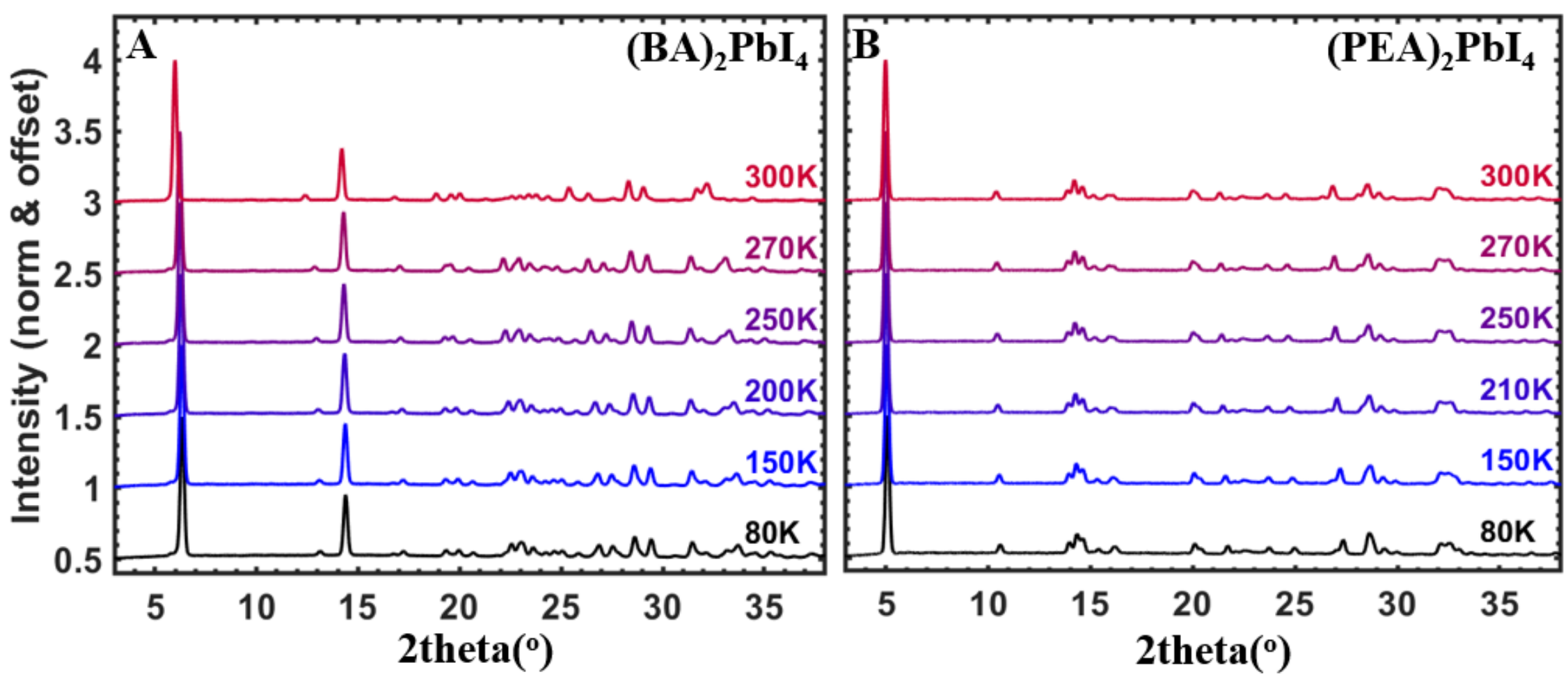

**FIG. 2** Temperature dependent X-ray diffraction patterns obtained from Bruker D8 VENTURE dual wavelength Mo/Cu Kappa four-circle diffractometer for (A)$(BA)_2PbI_4$ and (B)$(PEA)_2PbI_4$.

In these layered HOIPs, inorganic layer distortions are considered an important parameter that affects the structural and electronic properties. The distortions are generally manifested as octahedral tilts or re-alignment in these materials. Previous studies by Ziegler et al. [50] and Dyksik et al. [51] reported a decrease in the out-of-plane octahedral tilt angle in $(BA)_2PbI_4$ as the system transitions from the low-temperature (LT) phase (T $<$ 275 K) to the high-temperature (HT) phase (T $>$ 275 K). Similar observations were made in this work based on refining the temperature-dependent XRD patterns (Fig. 3). The octahedral tilt angle ($\delta$) was measured according to the schematic in Fig. 3(A) [51]. For $(BA)_2PbI_4$, $\delta$ does not show much change on warming from 80 K to 270 K (Fig. 3(B)), above which it drops from approximately 11.20° to 2.20° at the room temperature. This aligns with the structural phase transition at 275 K. For $(PEA)_2PbI_4$ in the temperature range of 80 K to 300 K, we couldn't find major changes in the average structure.

As reported by Gong and his colleagues [33], upon warming, PLQY remains almost constant up to 150 K for $(BA)_2PbBr_4$ and 225 K for $(PEA)_2PbBr_4$, and upon further warming it starts to decrease, which we think would be similar in $(BA)_2PbI_4$ and $(PEA)_2PbI_4$. Note that the study by Ziegler et al [50] indeed showed that for $(BA)_2PbI_4$, though octahedral tilts can be correlated to the electronic band gap, they do not have a dominant influence in the exciton binding energy, rather they remain more robust at the phase transition. Thus, the observed temperature-dependent evolution of octahedral tilts indicates that the distortion of inorganic layers does not correlate explicitly with PLQY in these two 2D HOIPs. A pictorial representation of these octahedral distortions in both $(BA)_2PbI_4$ and $(PEA)_2PbI_4$ is shown in Fig. 3(C) and (D).

Table. 1: Summary of crystal structure parameters based on the refinement of X-Ray diffraction patterns

| Compound | $\mathbf{(BA)_2PbI_4}$ | | $\mathbf{(PEA)_2PbI_4}$ | |
|---|---|---|---|---|
| **T/K** | 300 | 80 | 300 | 80 |
| Space group | $Pbca$ | $Pbca$ | $P\bar{1}$ | $P\bar{1}$ |
| **a**/Å | 8.8842(6) | 8.4145(3) | 8.762(1) | 8.687(3) |
| **b**/Å | 8.6963(5) | 8.9777(3) | 8.729(1) | 8.674(3) |
| **c**/Å | 27.6322(12) | 26.0739(7) | 33.017(11) | 32.336(10) |
| $\boldsymbol{\alpha}$° | 90 | 90 | 84.39(3) | 85.39(3) |
| $\boldsymbol{\beta}$/ ° | 90 | 90 | 84.79(1) | 85.15(2) |
| $\boldsymbol{\gamma}$° | 90 | 90 | 89.64(1) | 89.46(1) |

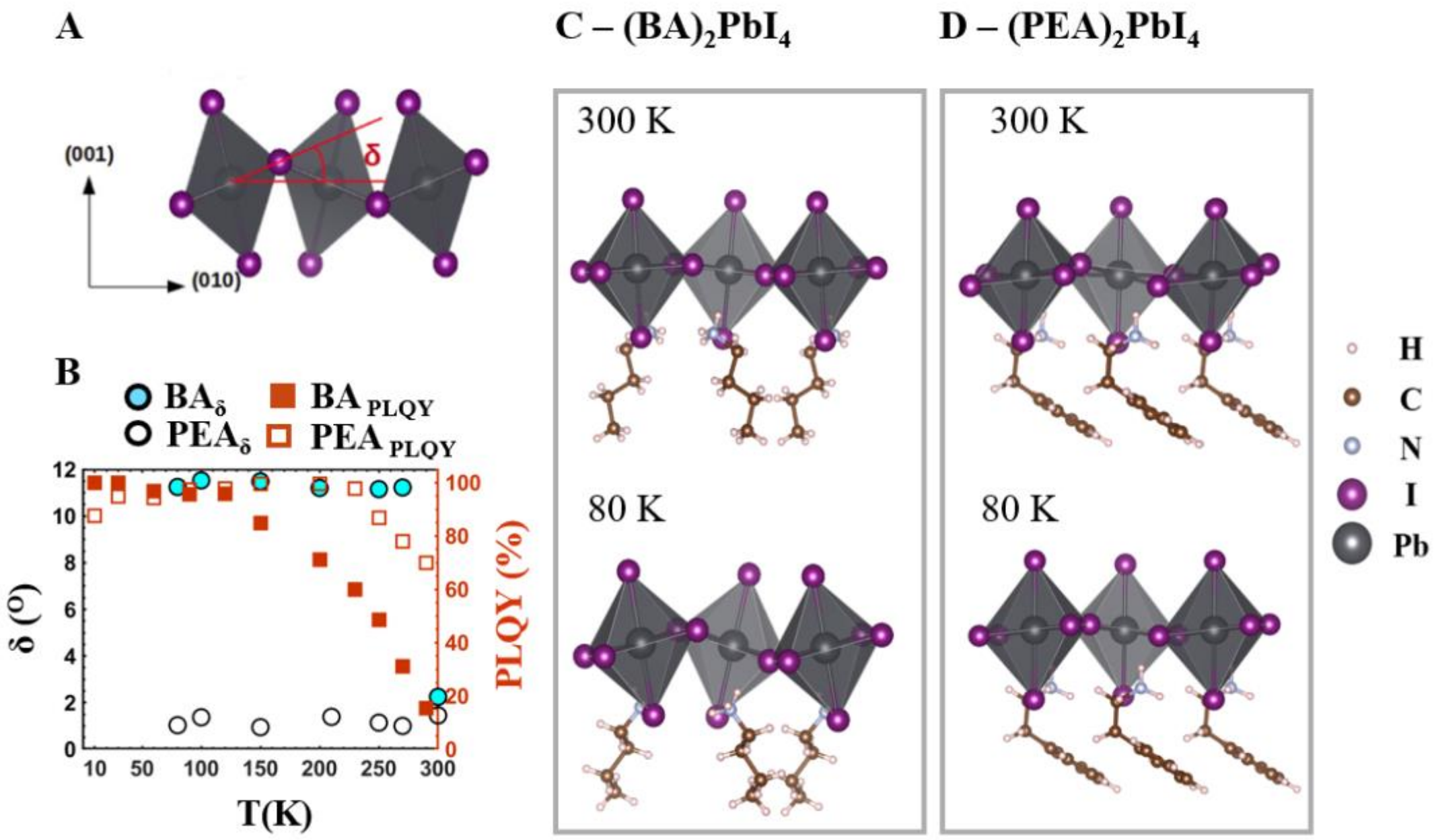

**FIG. 3** (A) Schematic of out of plane octahedral tilt angle (δ) referenced from [50]. (B) Comparison of variation of δ and PLQY with temperature for $(BA)_2PbI_4$ and $(PEA)_2PbI_4$. The black closed circles with cyan face color and black open circles represent the δ values for $(BA)_2PbI_4$ and $(PEA)_2PbI_4$ respectively. Orange closed and open squares represent the temperature dependent PLQY of $(BA)_2PbBr_4$ and $(PEA)_2PbBr_4$ respectively as reported in [33]. Schematic of octahedral distortion observed in (C) $(BA)_2PbI_4$ at 80K and 300K and (D) $(PEA)_2PbI_4$ at 80K and 300K. The images of crystal structures were generated using VESTA [43] after structural refinement on the XRD patterns using GSAS II [52].

### 3.2. Rotational dynamics

In a quantum well confinement structure, where the polarized organic molecule layer serves as the 'wall' and the inorganic layer serves as the 'well', the dynamics behaviors of both the 'wall' and 'well' are expected to dominate exciton dynamics. Inspired by influences from molecular re-orientations in 3D HOIPs [19], we first put our concentrations on the rotational dynamics of polarized organic molecules in these 2D HOIPs.

Due to neutrons' high sensitivity to hydrogen, the rotational motion of hydrogen-rich molecules can be directly probed as incoherent diffuse intensity, using Quasi Elastic Neutron Scattering (QENS) (Fig. 4). At ~ 180 K, the neutron scattering intensity concentrates sharply in the elastic channel for both systems. Upon heating, in $(BA)_2PbI_4$ the intensity quickly diffuses into the quasi-elastic region while in $(PEA)_2PbI_4$ the intensity mostly stays in the elastic channel and shows only slight increments in the quasi-elastic region. The enhanced QENS intensity in $(BA)_2PbI_4$ indicates much stronger molecular rotational dynamics in $(BA)_2PbI_4$ than in $(PEA)_2PbI_4$. Note that in both materials, the number of molecules in one unit cell is two and the number of hydrogen atoms on each molecule is the same, which means the incoherent QENS intensity is a direct manifestation of the robustness of molecular rotations.

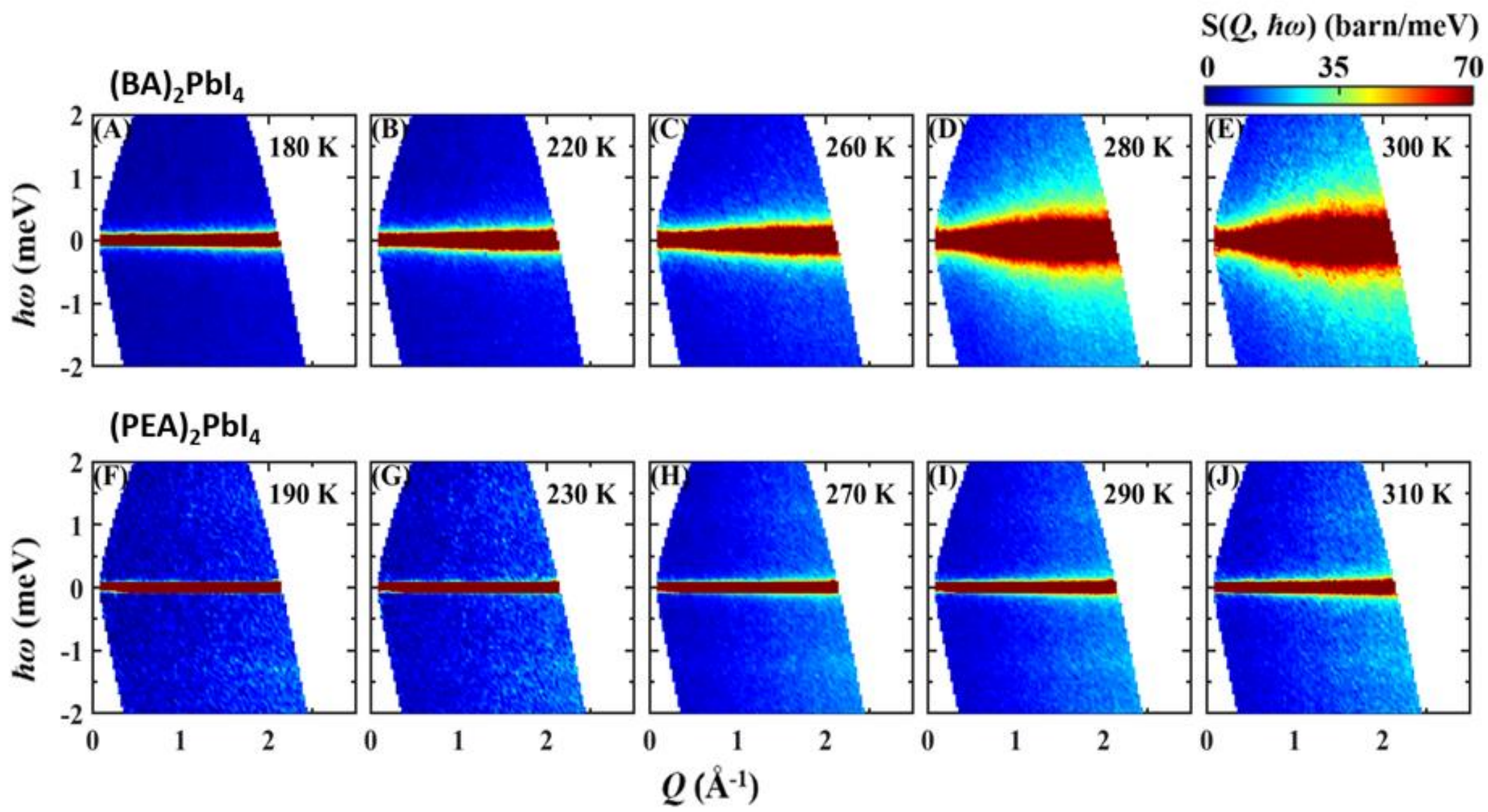

**FIG. 4** Temperature-dependent quasi-elastic neutron scattering spectra of $(BA)_2PbI_4$ (A-E) and $(PEA)_2PbI_4$ (F-J) obtained from AMATERAS upon heating.

To quantitatively understand the rotational motion of organic molecules, we applied the Jump model analysis [34, 35] along with group theory analysis. The rotation model that accounts for the preferential molecular orientation is called the Jump model [34, 35]. The rotational dynamics of the organic cation are determined by its symmetry and the local crystal symmetry. The possible rotational modes are described as the irreducible representations of the direct product $\Gamma = C \otimes M$ where $C$ and $M$ are the point groups of the local crystal symmetry and the molecule symmetry. In group theory, the static and dynamic structure factor for rotational motions of molecules in a crystal can be written as [35]

$$S_{cal}(Q, \hbar\omega) = e^{-\langle u^2 \rangle Q^2} \left( \sum_{\gamma} A_{\gamma}(Q) \frac{1}{\pi} \frac{\omega_{\gamma}}{1 + \omega^2 \tau_{\gamma}^2} \right)$$

(Eq. 1)

where the sum over $\gamma$ runs over all the irreducible representations of the system group $\Gamma$ ($\Gamma_{\gamma}$); $e^{-\langle u^2 \rangle Q^2}$ is the Debye Waller factor, $\langle u^2 \rangle$ is the mean squared atomic displacement. For a polycrystalline sample, $A_{\gamma}(Q)$ is given by [35]

$$A_\gamma(Q) = \frac{l_\gamma}{g}\sum_\alpha\sum_\beta \chi_\gamma^{\alpha\beta}\sum_{C_\alpha}\sum_{M_\beta} j_0\big(Q\big|R - C_\alpha M_\beta R\big|\big)$$

(Eq. 2)

Here $g$ is the order of group $\Gamma$ and $l_\gamma$ is the dimensionality of $\Gamma_\gamma$. The sums over $\alpha$ and $\beta$ run over all the classes of $C$ and $M$, respectively, and the sums over $C_\alpha$ and $M_\beta$ run over all the rotations that belong to the crystal class, $\alpha$, and to the molecule class, $\beta$, respectively. The characters of $\Gamma_\gamma$, $\chi_\gamma^{\alpha\beta}$,are the products of the characters of $C_{\gamma C}$ and $M_{\gamma M}$; $\chi_\gamma^{\alpha\beta} = \chi_{\gamma C}^{\alpha}\chi_{\gamma M}^{\beta}$. $j_0(x)$ is the zeroth spherical Bessel function and, $\left|R - C_\alpha M_\beta R\right|$, is the distance between the initial atom position $R$ and final atom position $C_\alpha M_\beta R$, called the jump distance. The relaxation time for the $\Gamma_\gamma$ mode, $\tau_\gamma$, is written as [35]

$$\frac{1}{\tau_\gamma} = \sum_\alpha \frac{n_\alpha}{\tau_\alpha}\left(1 - \frac{\chi_\gamma^{\alpha e}}{\chi_\gamma^{Ee}}\right) + \sum_\beta \frac{n_\beta}{\tau_\beta}\left(1 - \frac{\chi_\gamma^{E\beta}}{\chi_\gamma^{Ee}}\right)$$

(Eq. 3)

where $n_\alpha$, $n_\beta$ are the number of symmetry rotations of the classes, $\alpha$ and $\beta$, respectively. $E$ and $e$ represent the identity operations of $C$ and $M$, respectively.

To fit the quasi-elastic neutron scattering spectra, the calculated $S_{cal}(Q,\hbar\omega)$ has to be convoluted with the instrument resolution. The phonon contributions are also estimated in the form of incoherent phonon scattering. Then we have the fitting function as

$$S(Q,\hbar\omega) = A_{rot}\int_{-\infty}^{\infty} S_{cal}(Q,\hbar\omega - \hbar\omega')S_{res}(\hbar\omega')d(\hbar\omega') + A_{vib}Q^2e^{-\langle u^2\rangle Q^2}$$

(Eq. 4)

where $S_{res}(\hbar\omega)$ is the instrument resolution function. $A_{rot}$ and $A_{vib}$ are the scaling factors for the rotational contributions and the vibrational contributions.

Based on our previous study on the Jump model analysis of $MAPbI_3$ [35], we propose two rotational modes for $(BA)_2PbI_4$: three-fold ($C_3$) and four-fold ($C_4$) modes. Table 2 shows the character tables for point group $C_3$ and $C_4$. The terminal $NH_3$ and $CH_3$ groups experience rotational mode $\Gamma = C_4 \otimes C_3$ and the rest of $CH_2$ groups experience $\Gamma = C_4$. Based on direct product rules in group theory and equivalent atomic position distributions (Fig. 5), we can calculate the corresponding $\tau_\gamma$ and $A_\gamma(Q)$ for them (Table 3, 4). Eventually, we will see if the proposed rotational model works well with our experimental data.

**Table. 2** Character tables for $C_3$ and $C_4$. The point group $C_3$ has two irreducible representations: one one-dimensional representation A, and one two-dimensional representation E. The point group $C_4$ has three irreducible representations: two one-dimensional representations A and B, and one two-dimensional representation E.

| $C_3$ group | $E$ | $2C_3$ |
|---|---|---|
| A | 1 | 1 |
| $E$ | 2 | -1 |

| $C_4$ group | $E$ | $2C_4$ | $C_2$ |
|---|---|---|---|
| A | 1 | 1 | 1 |
| B | 1 | -1 | 1 |
| E | 2 | 0 | -2 |

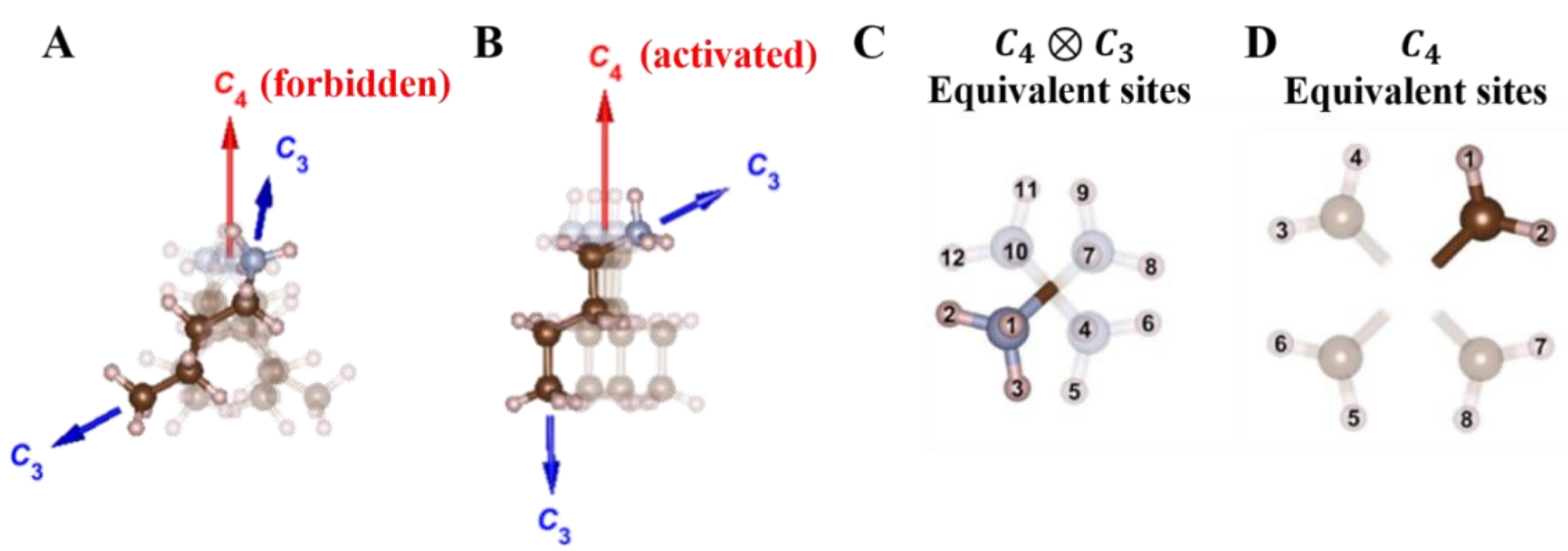

**FIG. 5** Rotational modes of $(BA)_2PbI_4$ and equivalent H sites for $NH_3$, $CH_3$, and $CH_2$ groups. (A and B) show the rotational modes of $BA^+$ molecules (A) When $C_4$ is frozen and terminal $NH_3$ and $CH_3$ groups experience $\Gamma = C_3$ (B) When the terminal $NH_3$ and $CH_3$ groups experience $\Gamma = C_4 \otimes C_3$ and the $CH_2$ groups experience $\Gamma = C_4$. (C) presents the 12 equivalent H sites for the $C_4 \otimes C_3$ mode of $NH_3$ and $CH_3$ groups. (D) presents the 8 equivalent H sites for the $C_4$ mode of $CH_2$ groups.

**Table. 3** Model details for jump mode $\Gamma = C_4 \otimes C_3$.

| $\Gamma_\gamma$ | $\frac{1}{\tau_\gamma}$ | $36 \cdot A_\gamma(Q)$ |
|---|---|---|
| A⊗A | 0 | $3 + 6j_1 + 2j_2 + 2j_3 + 2j_4 + j_5 + 2j_6 + 2j_7 + 2j_8 + 2j_9$ $+2j_{10} + 2j_{11} + j_{12} + 2j_{13} + 2j_{14} + 2j_{15} + j_{16}$ |
| A⊗E | $\frac{3}{\tau_{C_3}}$ | $6 - 6j_1 + 4j_2 - 2j_3 - 2j_4 + 2j_5 - 2j_6 - 2j_7 - 2j_8 - 2j_9$ $+4j_{10} - 2j_{11} + 2j_{12} - 2j_{13} - 2j_{14} + 4j_{15} + 2j_{16}$ |
| B⊗A | $\frac{4}{\tau_{C_4}}$ | $3 + 6j_1 - 2j_2 - 2j_3 - 2j_4 + j_5 + 2j_6 + 2j_7 - 2j_8 - 2j_9$ $-2j_{10} - 2j_{11} + j_{12} + 2j_{13} - 2j_{14} - 2j_{15} + j_{16}$ |
| B⊗E | $\frac{4}{\tau_{C_4}} + \frac{3}{\tau_{C_3}}$ | $6 - 6j_1 - 4j_2 + 2j_3 + 2j_4 + 2j_5 - 2j_6 - 2j_7 + 2j_8 + 2j_9$ $-4j_{10} + 2j_{11} + 2j_{12} - 2j_{13} + 2j_{14} - 4j_{15} + 2j_{16}$ |
| E⊗$A$ | $\frac{2}{\tau_{C_4}}$ | $6 + 12j_1 - 2j_5 - 4j_6 - 4j_7 - 2j_{12} - 4j_{13} - 2j_{16}$ |
| $E$⊗E | $\frac{2}{\tau_{C_4}} + \frac{3}{\tau_{C_3}}$ | $12 - 12j_1 - 4j_5 + 4j_6 + 4j_7 - 4j_{12} + 4j_{13} - 4j_{16}$ |

Here $j_i$ represent the zeroth spherical Bessel function $j_0(Qr_i)$, where $r_i$ are the jump distances corresponding to the jump positions in Fig. 5: $r_1 = R_{1,2}$, $r_2 = R_{1,4}$, $r_3 = R_{1,5}$, $r_4 = R_{1,6}$, $r_5 = R_{1,7}$, $r_6 = R_{1,8}$, $r_7 = R_{1,9}$, $r_8 = R_{1,11}$, $r_9 = R_{1,12}$, $r_{10} = R_{2,5}$, $r_{11} = R_{2,6}$, $r_{12} = R_{2,8}$, $r_{13} = R_{2,9}$, $r_{14} = R_{2,12}$, $r_{15} = R_{3,6}$, $r_{16} = R_{3,9}$. $R_{i,j} = |\mathbf{R_i} - \mathbf{R_j}|$, where $\mathbf{R_i}$ is the position of the i-th H site. The 12 equivalent H sites for $NH_3$ and $CH_3$ are marked in Fig. 5 (C).

**Table. 4** Model details for jump mode $\Gamma = C_4$.

| $\Gamma_\gamma$ | $\frac{1}{\tau_\gamma}$ | $8 \cdot A_\gamma(Q)$ |
|---|---|---|
| A | 0 | $2 + 2j_2 + j_4 + 2j_7 + j_8$ |
| B | $\frac{4}{\tau_{C_4}}$ | $2-2j_2+j_4-2j_7+j_8$ |
| E | $\frac{2}{\tau_{C_4}}$ | $4-2j_8$ |

Here $j_i$ represents the zeroth spherical Bessel function $j_0(Qr_i)$, where $r_i$ are the jump distances corresponding to the jump positions in Fig. 5: $r_2 = R_{1,3}$, $r_4 = R_{1,5}$, $r_7 = R_{2,4}$, $r_8 = R_{2,6}$. $R_{i,j} = |\mathbf{R_i} - \mathbf{R_j}|$, where $\mathbf{R_i}$ is the position of the i-th H site. The 8 equivalent H sites for $CH_2$ groups are marked in Fig. 5 (D).

For $(PEA)_2PbI_4$, only one rotational mode is proposed: the $C_3$ mode of the terminal $NH_3$ group (Table. 2). The structure factors $A_\gamma(Q)$ for $\Gamma = C_3$ is calculated in Table. 5.

**Table. 5** Model details for jump mode $\Gamma = C_3$

| $\Gamma_\gamma$ | $\frac{1}{\tau_\gamma}$ | $9 \cdot A_\gamma(Q)$ |
|---|---|---|
| A | 0 | $3+6j_0(Qr)$ |
| E | $\frac{3}{\tau_{C_3}}$ | $6-6j_0(Qr)$ |

$j_0(Qr)$ is the zeroth spherical Bessel function. r is the jump distance between H atoms of the $NH_3$ group.

The proposed rotational models for the two samples work well with the experimental data (Fig. 6 and Fig. 7), demonstrating their validity. The fitted parameters are reported in Table. 6. The fitting based on proposed models suggests that there are two intrinsic rotational modes for $(BA)_2PbI_4$ : A $C_3$ mode of the terminal $NH_3$ and $CH_3$ groups in BA molecule at both low-temperature (LT) $Pbca$ phase ( $T < 275$ K ) and high-temperature (HT) Pbca phase ($T > 275$ K ), and a $C_4$ mode of the entire BA molecule about the crystallographic c-axis which gets activated only at the HT Pbca phase. For $(PEA)_2PbI_4$, our analysis suggests a single $C_3$ mode of the terminal $NH_3$ group throughout the temperature of interest. At the base temperature for both systems (10K for $(BA)_2PbI_4$ and 8K for $(PEA)_2PbI_4$) any rotations are frozen so that the elastic channel intensity of QENS data can be described just with the instrument resolution function ($S_{res}(\hbar\omega)$).

In 3D HOIPs, our previous study [53] showed that the entropy contribution to the Gibbs free energy caused by rotations of polarized organic molecules plays a significant role in structural phase transition. Here in these 2D HOIPs, the BA molecule possesses larger rotational moments of inertia than the PEA molecule, which brings it higher rotational entropy. Therefore, it is reasonable to expect that molecular rotations would have a greater impact on the dynamics of the inorganic framework and the dielectric environment surrounding confined excitons in $(BA)_2PbI_4$ compared to $(PEA)_2PbI_4$. It is important to note that the elastic channel intensity ($-0.05 < \hbar\omega < 0.05$ meV) decays much faster in Q, for $(BA)_2PbI_4$ above 180 K (Fig. 6(A)), while for $(PEA)_2PbI_4$ (Fig. 7(A)), it do not show a significant change. This is reflected in the lifetimes of the C3 rotational modes for both systems (Fig. 8). Comparing similar temperatures, the C3 rotational modes for $(BA)_2PbI_4$ has shorter rotational relaxation times than $(PEA)_2PbI_4$. This suggests more robust molecular rotational motions in $(BA)_2PbI_4$. These observations imply that the rotational dynamics of organic molecules in these 2D HOIPs could significantly influence charge dynamics and optoelectronic performance. To understand the affects of molecular rotations, we compare the temperature-dependent evolution of QENS spectra with that of PLQY, which will be discussed in later sections.

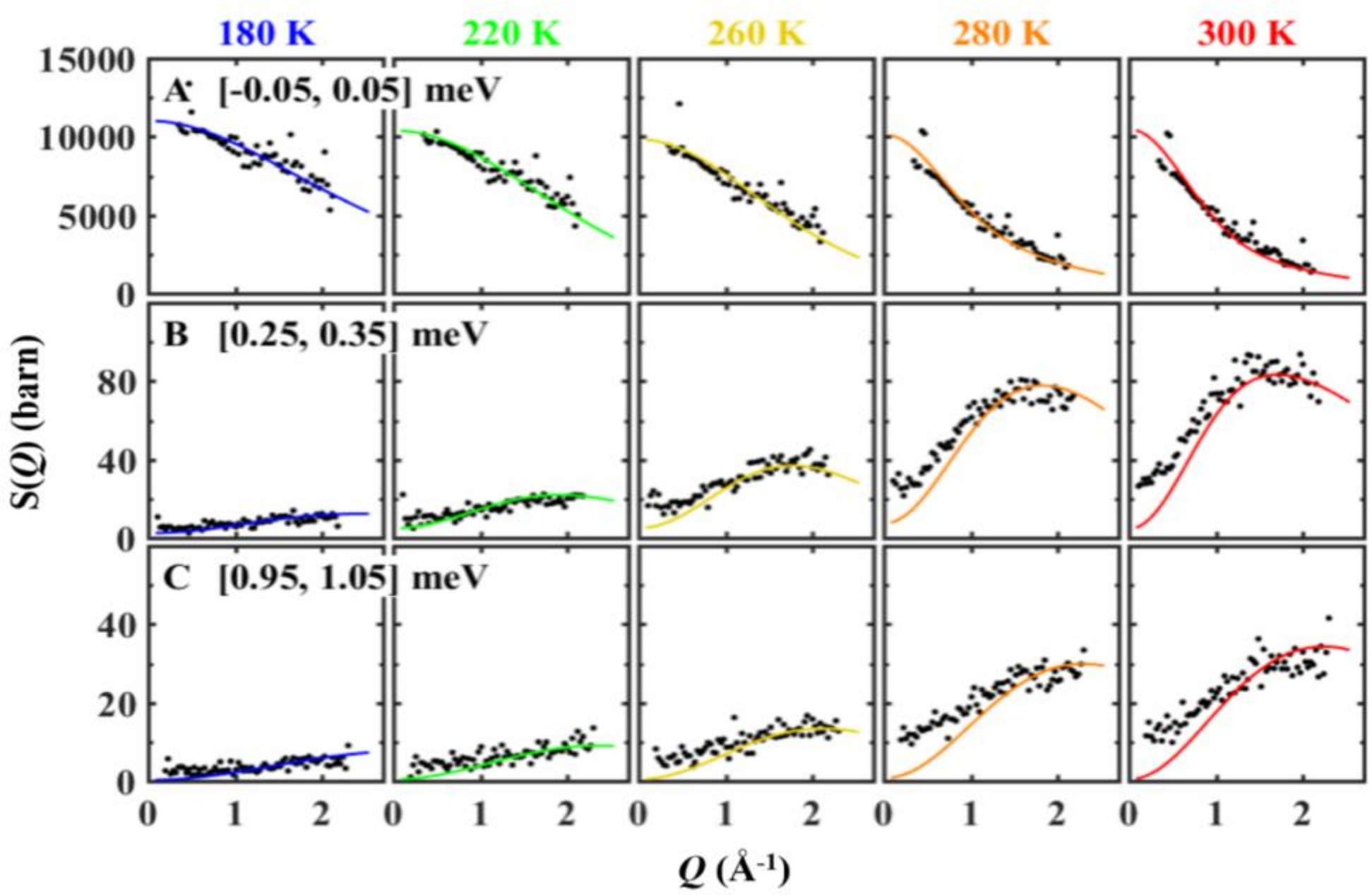

**FIG. 6** Constant energy slices of QENS spectra and fitting results of $(BA)_2PbI_4$. (A - C) the $\hbar\omega$-integrated QENS data, S(Q), over three different energy ranges, $-0.05 < \hbar\omega < 0.05$ meV (A), $0.25 < \hbar\omega < 0.35$ meV (B), $0.95 < \hbar\omega < 1.05$ meV (C), with five selected temperatures, 180 K, 220 K, 260 K (low-T *Pbca* orthorhombic phase), 280 K, 300 K (high-T *Pbca* orthorhombic phase). The black dots are the measured data, and the colored solid lines are the model-fitted QENS intensity.

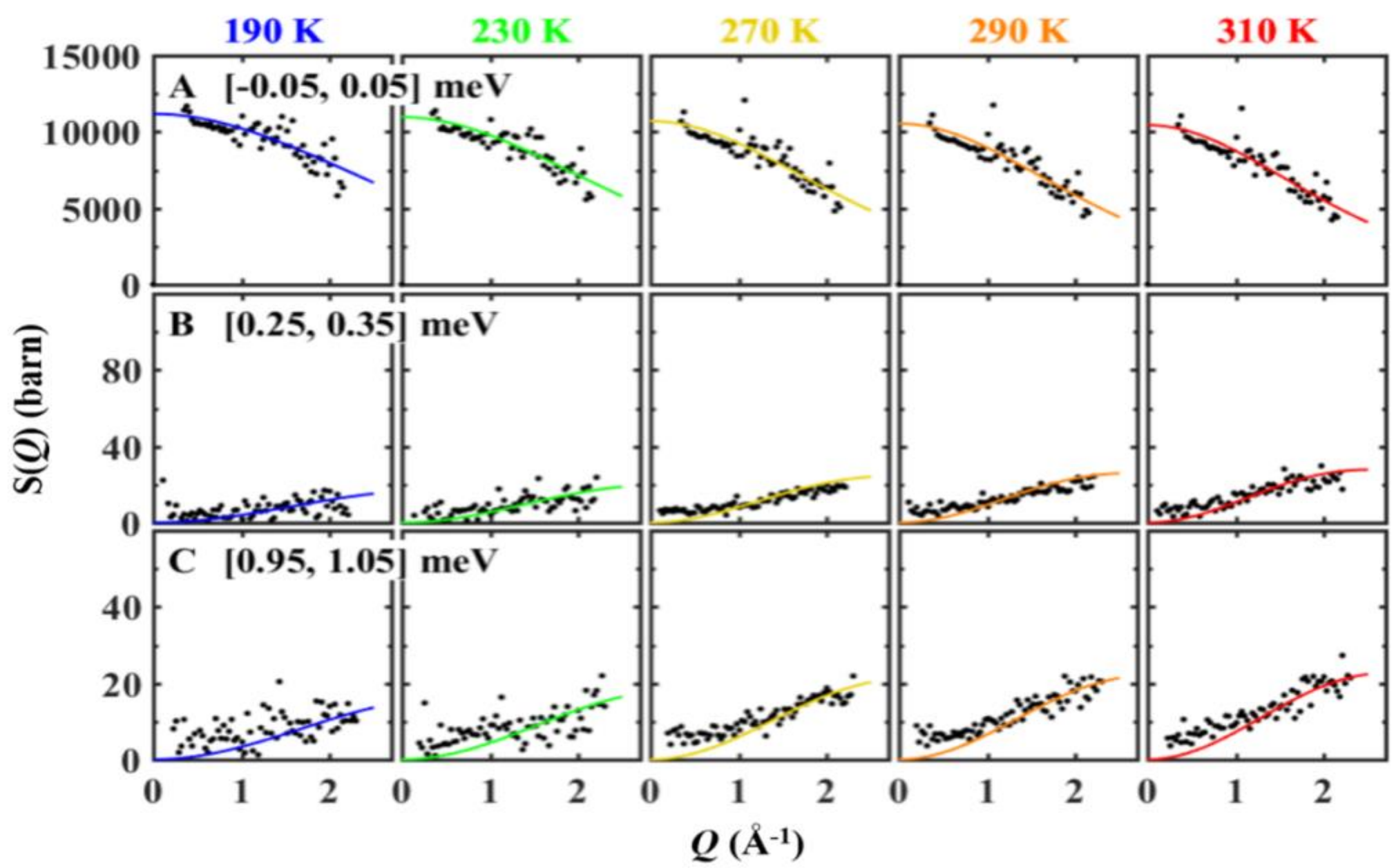

**FIG. 7** Constant energy slices of QENS spectra and fitting results of $(PEA)_2PbI_4$. (A - C) the $\hbar\omega$-integrated QENS data, S(Q), over three different energy ranges, $-0.05 < \hbar\omega < 0.05$ meV (A), $0.25 < \hbar\omega < 0.35$ meV (B), $0.95 < \hbar\omega < 1.05$ meV (C), with five selected temperatures, 190 K, 230 K, 270 K, 290 K, 310 K (triclinic *P-1* single phase). The black dots are the measured data, and the colored solid lines are the model-fitted QENS intensity.

**Table. 6** Estimated relaxation times, $\tau_{C_4}$ and $\tau_{C_3}$, and the mean squared displacement for the rotations of organic molecules in $(BA)_2PbI_4$ and $(PEA)_2PbI_4$ that are extracted from the model fitting to the QENS data as discussed in the text. Values in the parentheses indicate their errors.

| $(BA)_2PbI_4$ | | | | $(PEA)_2PbI_4$ | | |
|---|---|---|---|---|---|---|
| $T$ (K) | $\tau_{C_4}$ (ps) | $\tau_{C_3}$ (ps) | $\langle u^2\rangle$ (Å$^2$) | $T$ (K) | $\tau_{C_3}$ (ps) | $\langle u^2\rangle$ (Å$^2$) |
| 160 | ∞ | 342(10) | 0.110(4) | 170 | 588(90) | 0.067(4) |
| 180 | ∞ | 201(4) | 0.108(3) | 190 | 497(66) | 0.073(4) |
| 200 | ∞ | 104(2) | 0.116(3) | 210 | 408(50) | 0.079(4) |
| 220 | ∞ | 87(2) | 0.104(3) | 230 | 243(20) | 0.087(4) |
| 240 | ∞ | 70(2) | 0.122(4) | 250 | 177(12) | 0.095(4) |
| 260 | ∞ | 45(1) | 0.156(3) | 270 | 125(7) | 0.103(3) |
| 280 | 60(3) | 6.8(2) | 0.085(2) | 290 | 92(5) | 0.110(3) |
| 300 | 25(2) | 3.5(2) | 0.082(3) | 310 | 71(4) | 0.118(3) |

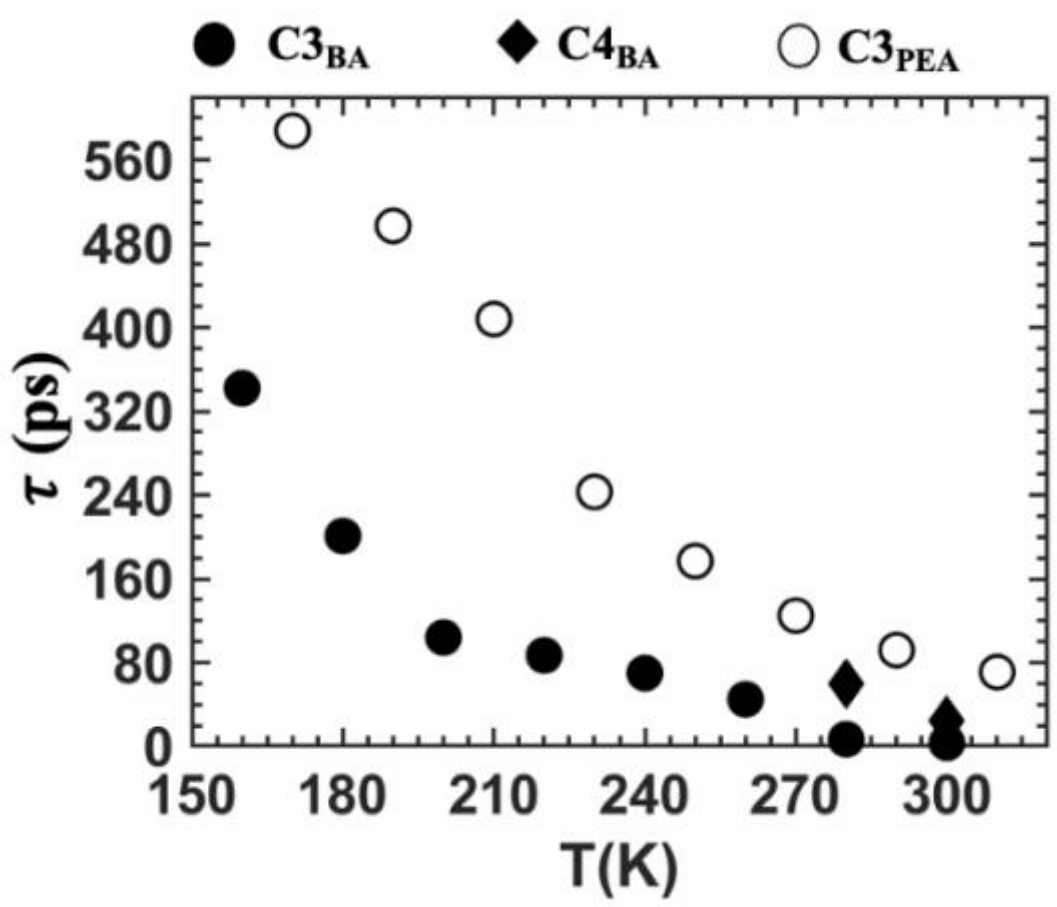

**FIG. 8** Estimated relaxation times ($\tau$) for the rotations of organic molecules in $(BA)_2PbI_4$ and $(PEA)_2PbI_4$ that are extracted from the model fitting to the QENS data. The closed circles represent three-fold ($C_3$) mode of the terminal $NH_3$ and $CH_3$ groups in BA molecule. The closed diamonds represent four-fold ($C_4$) mode along the crystallographic c-axis of BA molecule. The open circles represent the three-fold ($C_3$) mode of the terminal $NH_3$ group in PEA molecule.

### 3.3. Vibrational dynamics

To properly evaluate the potential influences from lattice vibrations, we need to carefully categorize the multiple types of phonons in these 2D HOIPs as they carry different functionalities in optoelectronic performance. Based on their refined crystal structures, we performed DFT calculations of their phonon band structures and simulated base temperature Inelastic Neutron Scattering (INS) spectra using OCLIMAX [54]. The simulated INS spectra of $(BA)_2PbI_4$ reproduced the observed data decently well (Fig. 9 and Fig. 11 (A)), but for $(PEA)_2PbI_4$ there are some discrepancies in the low energy ($\hbar\omega < 10$meV) region (Fig. 10 and Fig. 11(B)). This could be due to the low crystal symmetry (triclinic $P\bar{1}$) of $(PEA)_2PbI_4$ with many 188 atoms in one crystal unit cell, which brings in structural instability and makes the calculation of low-energy acoustic and optical phonons much more challenging.

Based on the DFT calculations, we characterized the phonon modes into three different types: inorganic phonon modes, hybrid phonon modes, and organic phonon modes based on the vibration energy fraction (VEF) calculation. VEF determines the fractional energy contribution of each atomic type in the unit cell, for the phonon modes present (see supplementary information section 1 for VEF calculation details). Figures 9 (D) and 10 (D), show the VEF for $(BA)_2PbI_4$ and $(PEA)_2PbI_4$ respectively. The phonon modes are classified as inorganic when the majority of atomic vibrations originate from inorganic atoms, Pb and I, with $VEF_{Pb,I} > 60 \sim 70\%$. Hybrid phonons have significant contributions from both inorganic and organic atoms (C, H, N) with $2\% < VEF_{Pb,I} < 60\%$ and organic phonons with dominant contributions from organic atoms with $VEF_{Pb,I} < 2\%$. The inorganic phonons mostly reside below $\hbar\omega$ < 10 meV, the hybrid phonons between 10 meV < $\hbar\omega$ < 35 meV, and the pure organic phonons above $\hbar\omega$ >~ 35 meV. The inorganic and hybrid phonons are expected to affect the optoelectronic properties of ionic crystals since they could change the highly polarized ionic bonds inside the samples, which can influence the charge carrier dynamics via Coulomb interactions. Whereas, the pure organic phonons, which involve molecular internal vibration and covalence bond deformation, play a minor role in the optoelectronic performance of the material as they barely show significant temperature dependence (Fig. 12). For $(BA)_2PbI_4$ above 35 meV, the scattering intensity do not show much change with temperature (Fig. 12(A)). For $(PEA)_2PbI_4$ other than a shift in scattering intensity from base temperature (8 K) to higher temperatures, which we attribute to the increased thermal population, we do not see major temperature dependence above 35 meV (Fig. 12(B)). Therefore, we focused on evaluating inorganic and hybrid phonons.

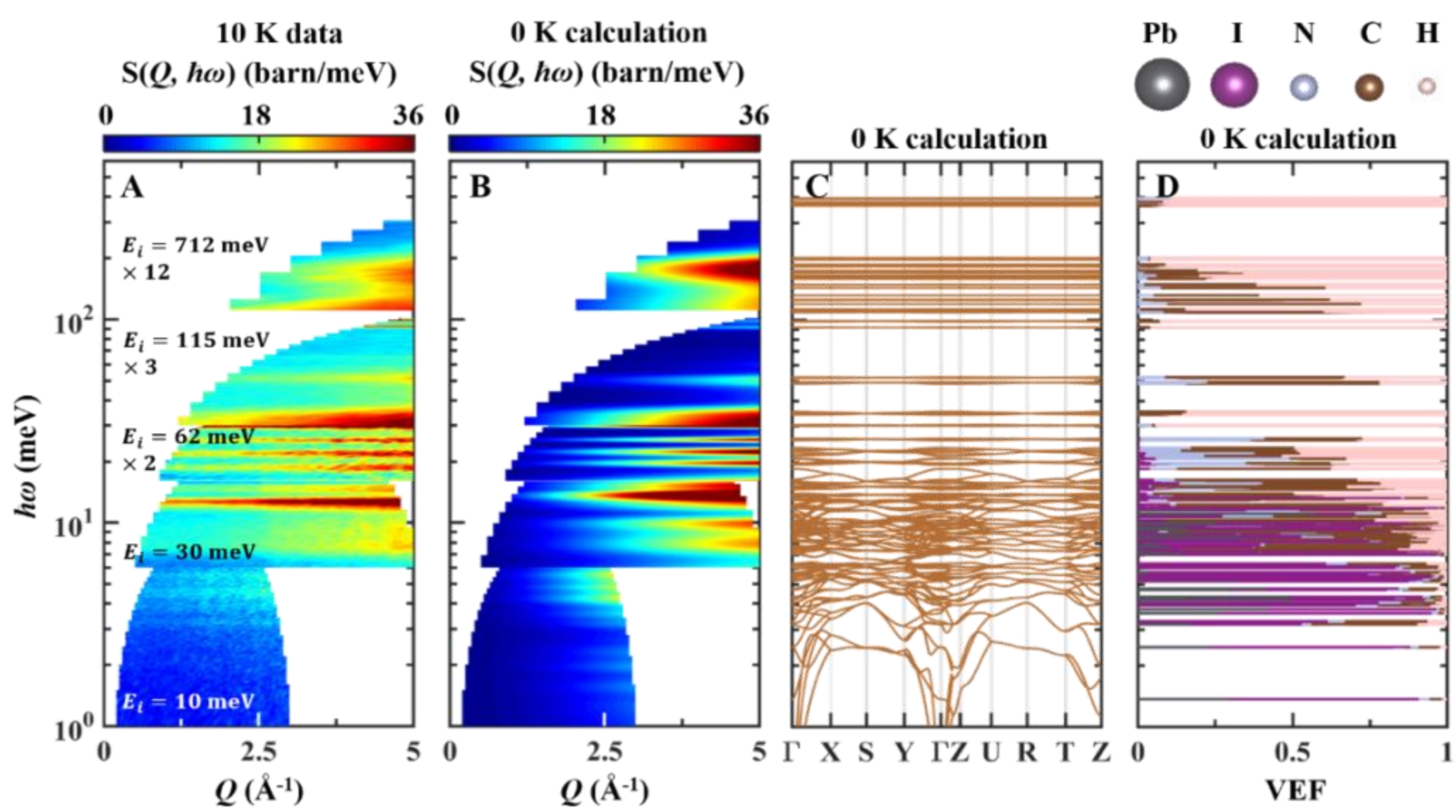

**FIG. 9** Experimental phonon spectra and DFT calculation results of $(BA)_2PbI_4$. (A) shows the experimental phonon spectra of $(BA)_2PbI_4$ taken at 10 K with $E_i = 10, 30, 62, 115, 712$ meV. The intensities of $E_i = 62$, 115, and 712 meV are rescaled by 2, 3, and 12 respectively. (B) shows the simulated phonon spectra using software OCLIMAX [54]. (C) presents the calculated phonon band structure along high-symmetry reciprocal **Q** points, $\Gamma = (0,0,0)$, $X = \left(\frac{1}{2},0,0\right)$, $S = \left(\frac{1}{2},\frac{1}{2},0\right)$, $Y = \left(0,\frac{1}{2},0\right)$, $Z = \left(0,0,\frac{1}{2}\right)$, $U = \left(\frac{1}{2},0,\frac{1}{2}\right)$, $R = \left(\frac{1}{2},\frac{1}{2},\frac{1}{2}\right)$, and $T = \left(0,\frac{1}{2},\frac{1}{2}\right)$. (D) contains the vibrational energy fractions at the Γ point for each phonon mode of $(BA)_2PbI_4$. Here the gray, violet, cyan, brown, and pink spheres represent the energy fractions of Pb, I, N, C, and H atoms, respectively.

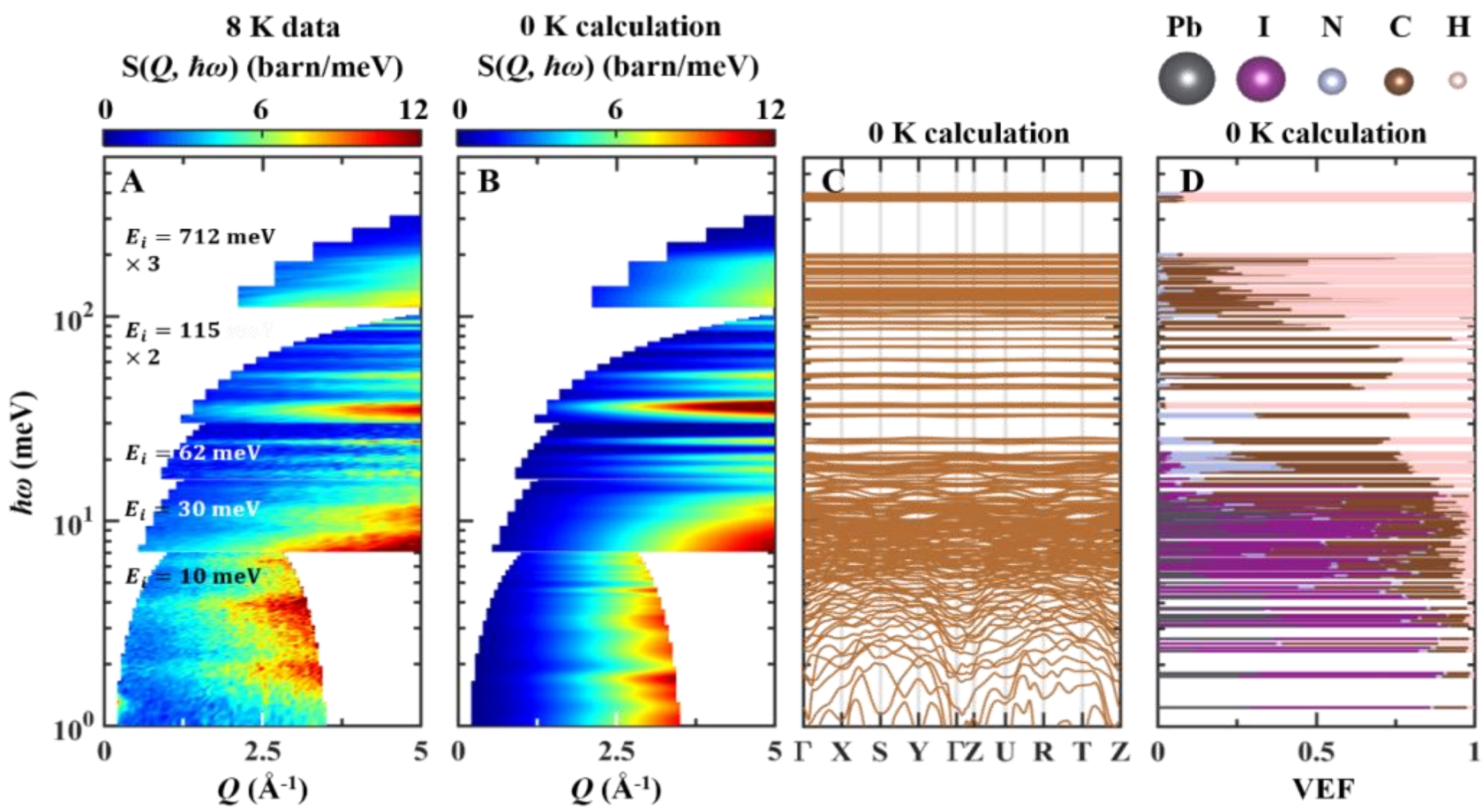

**FIG. 10** Experimental phonon spectra and DFT calculation results of $(PEA)_2PbI_4$. (A) shows the experimental phonon spectra of $(PEA)_2PbI_4$ taken at 8 K with $E_i = 10, 30, 62, 115, 712$ meV. The intensities of $E_i = 115$ and 712 meV are rescaled by 2 and 3 respectively. (B) shows the simulated phonon spectra using software OCLIMAX [54]. (C) presents the calculated phonon band structure along high-symmetry reciprocal **Q** points, $\Gamma = (0,0,0)$, $X = \left(\frac{1}{2},0,0\right)$, $S = \left(\frac{1}{2},\frac{1}{2},0\right)$, $Y = \left(0,\frac{1}{2},0\right)$, $Z = \left(0,0,\frac{1}{2}\right)$, $U = \left(\frac{1}{2},0,\frac{1}{2}\right)$, $R = \left(\frac{1}{2},\frac{1}{2},\frac{1}{2}\right)$, and $T = \left(0,\frac{1}{2},\frac{1}{2}\right)$. (D) contains the vibrational energy fractions at the Γ point for each phonon mode of $(PEA)_2PbI_4$. Here the gray, violet, cyan, brown, and pink spheres represent the energy fractions of Pb, I, N, C, and H atoms, respectively.

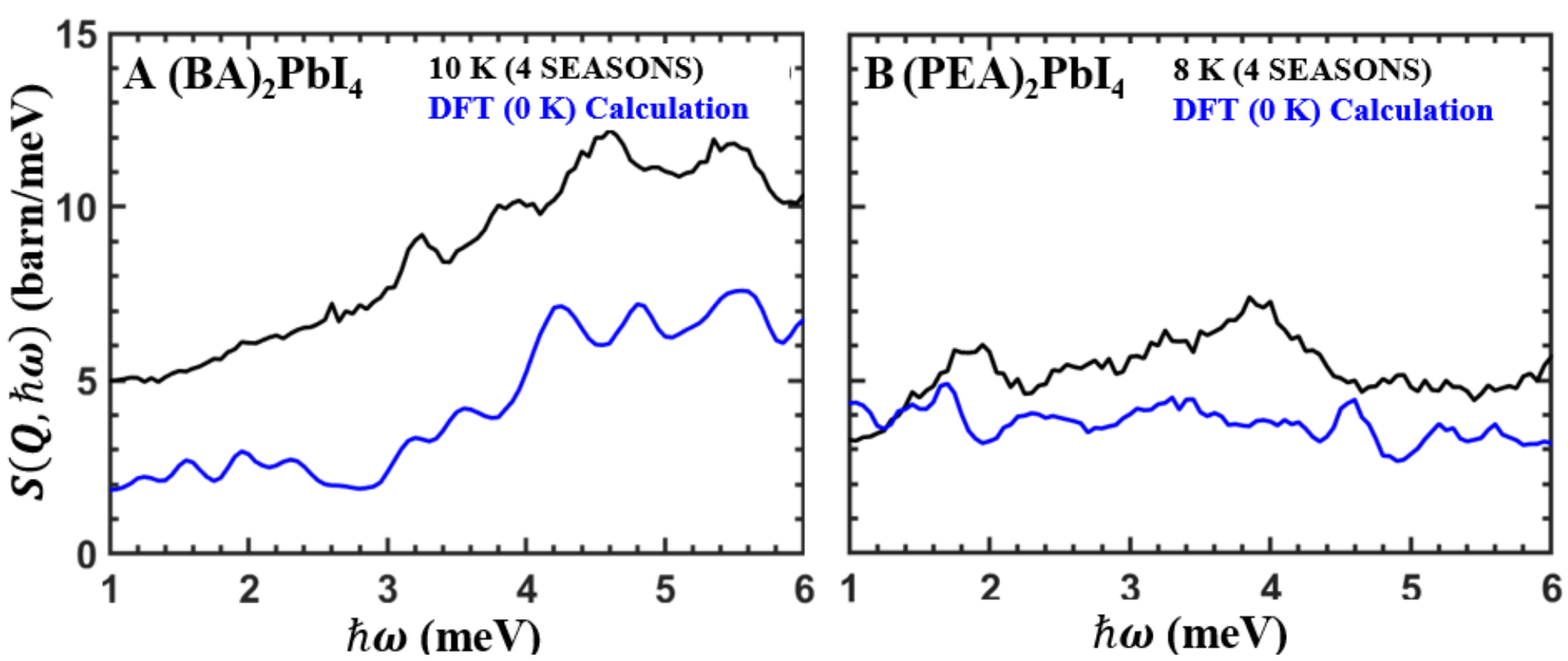

**FIG. 11** Q-integrated phonon spectra obtained from 4 SEASONS for $(BA)_2PbI_4$( A) and $(PEA)_2PbI_4$( B). The Q-integration range is selected as $[1.5 - 2]\text{Å}^{-1}$ which covers dominant phonon signal for both systems.

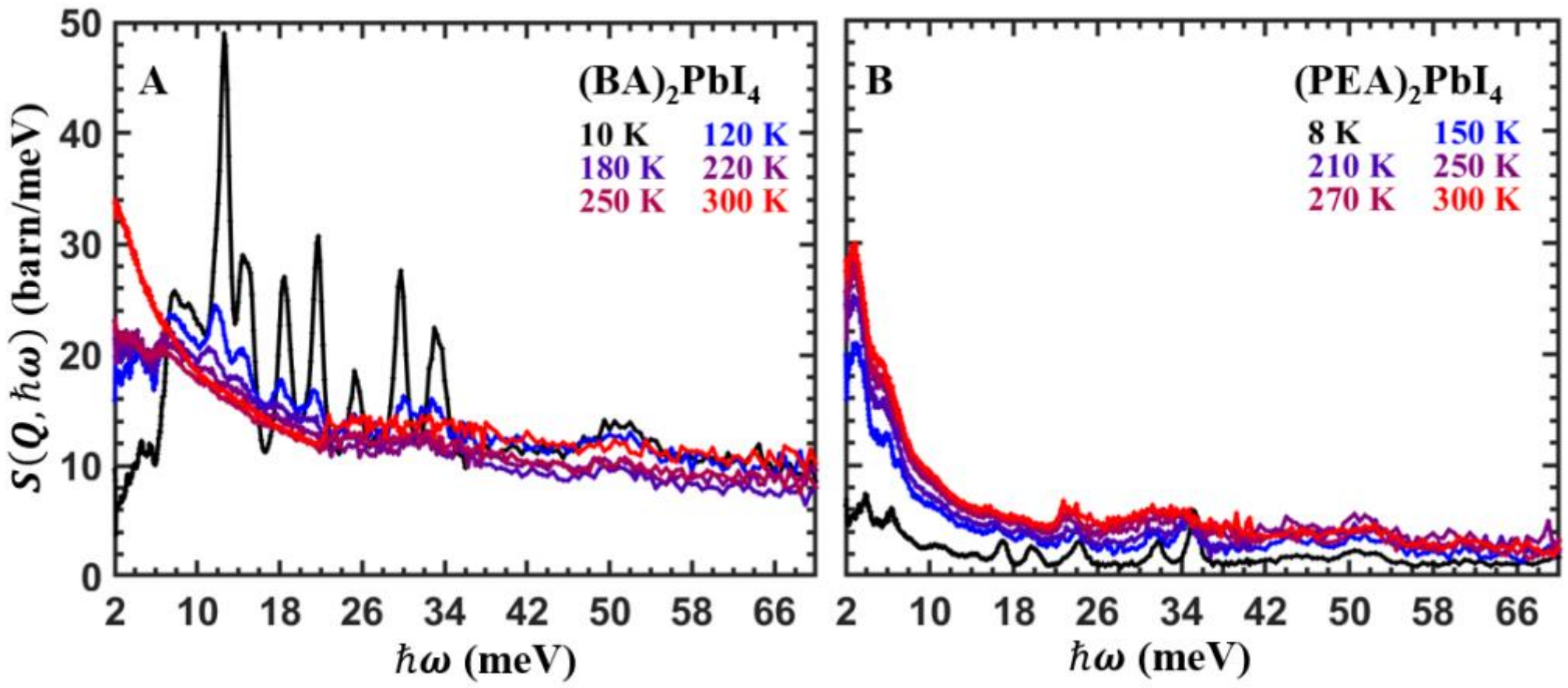

**FIG. 12** Q-integrated phonon spectra after connecting the QENS data from AMATERAS and INS data from 4SEASONS.at ~2 meV, for $(BA)_2PbI_4$( A) and $(PEA)_2PbI_4$( B). The Q-integration range is selected as $[1.5 - 2]Å^{-1}$ which covers dominant phonon signal for both systems. The pure inorganic phonons ($\hbar\omega \leq 10.0$meV), and the hybrid phonons ($10.0 \leq \hbar\omega \leq 35.0$meV) showed significant temperature dependence while the scattering intensities from pure organic phonons ($\hbar\omega \geq 35.0$meV) remained approximately constant across all temperatures.

For a comprehensive analysis, combining neutron scattering spectra from different incident energies is crucial to ensure a continuous dataset across all energy transfers. This requires normalizing the measured neutron scattering intensities to absolute units. Due to the large number of hydrogen atoms, the incoherent scattering background is prominent in both systems. Thus, we normalized the neutron scattering using elastic incoherent scattering intensities [55]. After normalization, we connected the QENS spectra collected on AMATERAS to the INS spectra from 4SEASONS at $\hbar\omega \sim 2.25$ meV (Figure 13(A)(B)). To separate the rotational intensity contributions from vibrational contributions, we applied a single Voigt function centered at $\hbar\omega = 0$ with elastic energy resolution convoluted to represent the molecular rotational contribution and multiple inelastic Voigt functions representing phonon peaks to interpret the rest of INS spectra (Fig. 13(C-F)). Upon warming, the phonon peaks become ill-defined and almost smear out at 300 K, which makes it difficult to identify. By fixing the number of phonon peaks and limiting their energy shifts, we managed to reproduce the phonon damping process (peak broadening), although this approach posed significant challenges. In addition, in $(BA)_2PbI_4$ the hybrid phonon intensity counter-intuitively decreased upon warming (Fig. 13(A)).

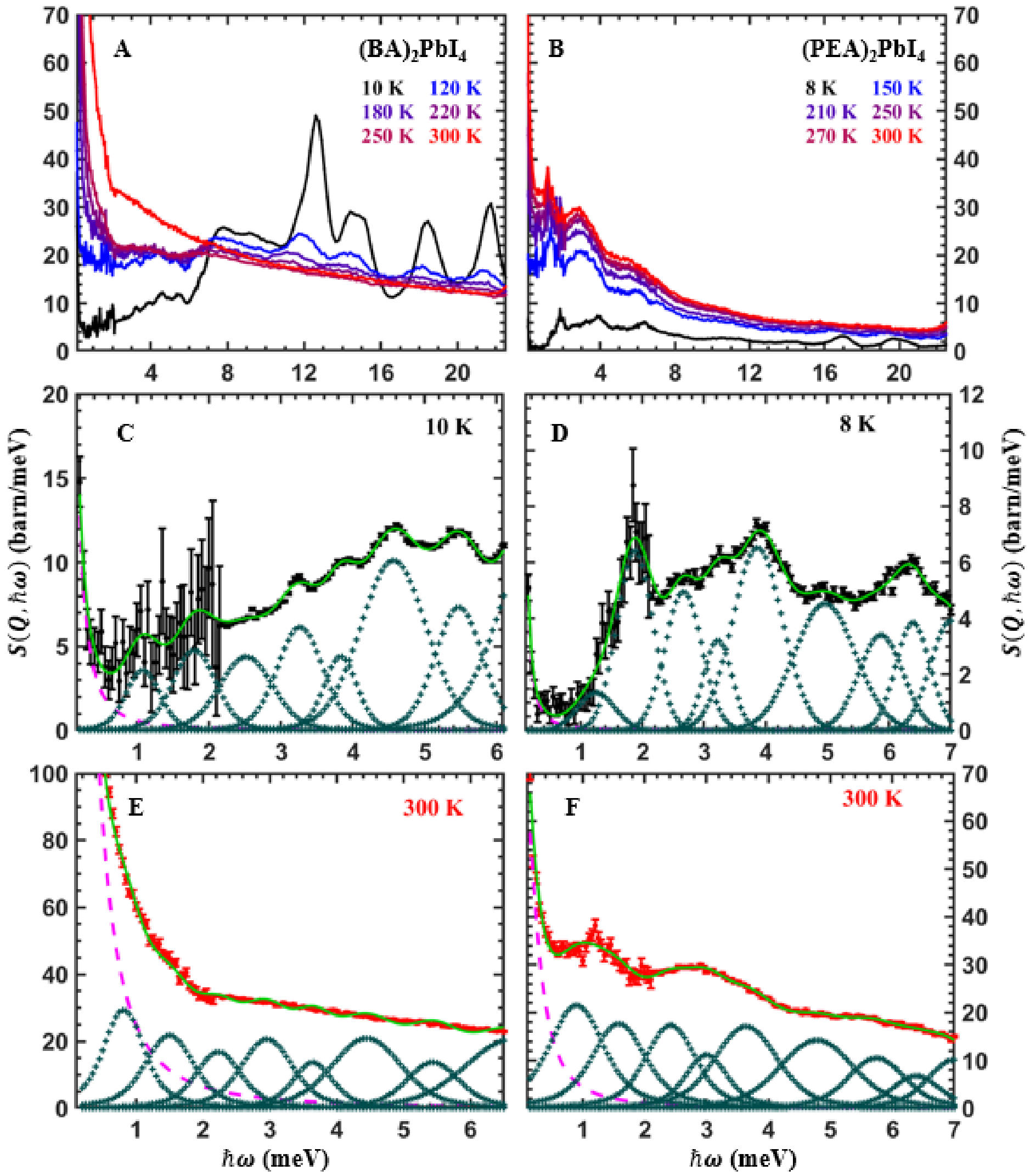

**FIG. 13 Q**-integrated phonon spectra of $(BA)_2PbI_4$ (A) and $(PEA)_2PbI_4$ (B) as a function of temperature. The above phonon spectra is plotted after connecting the AMATERAS and 4SEASONS data at 2.25 meV. The Q-integration range is selected as [1.5-2] $Å^{-1}$ which covers dominant phonon signal for both systems. Panel C, E and D, F shows the rotational and vibrational contributions after fitting for $(BA)_2PbI_4$ and $(PEA)_2PbI_4$ respectively. The rotational contributions are represented as the magenta dashed lines and the vibrational contributions as the (+) dark green symbols.

To back up our analysis, we performed temperature-dependent Raman scattering measurements on powder samples of the two materials (Fig. 14). In Raman scattering the Raman inactive rotational modes would be 'filtered' out, leaving the dominant features arising from lattice vibrations. In other words, Raman spectroscopy could help better analyze low energy lattice vibration spectra, which would be 'contaminated' in neutron scattering by the huge incoherent scattering cross section of hydrogen atoms. On the other hand, due to the same factor, the rotational dynamics of hydrogen-rich molecules could be well analyzed in QENS spectra. Comparing the inelastic neutron scattering spectra to Raman spectra for similar temperatures, for both $(BA)_2PbI_4$ (Fig. 14(A)) and $(PEA)_2PbI_4$ (Fig. 14(B)), the phonon peaks are better defined in the Raman spectra. This is because of the difference in the scattering techniques. In Raman scattering, the wavelength of the incident visible light of 785 nm is much larger than the interatomic distances. Hence the Raman scattering can probe only the phonons with wavevectors (momentum transfers) $\lesssim 8 \times 10^{-4}$ Å$^{-1}$, in other words, the Raman scattering probes phonons at Γ point (Q = 0) [56], while our neutron scattering probes the powder averaged contribution as a function of Q. Additionally, both the inorganic and hybrid phonon intensities in the Raman spectra intuitively get enhanced upon warming along with graduate peak broadening. All these inelastic spectra will be used in later comparison with the temperature-dependent PLQY.

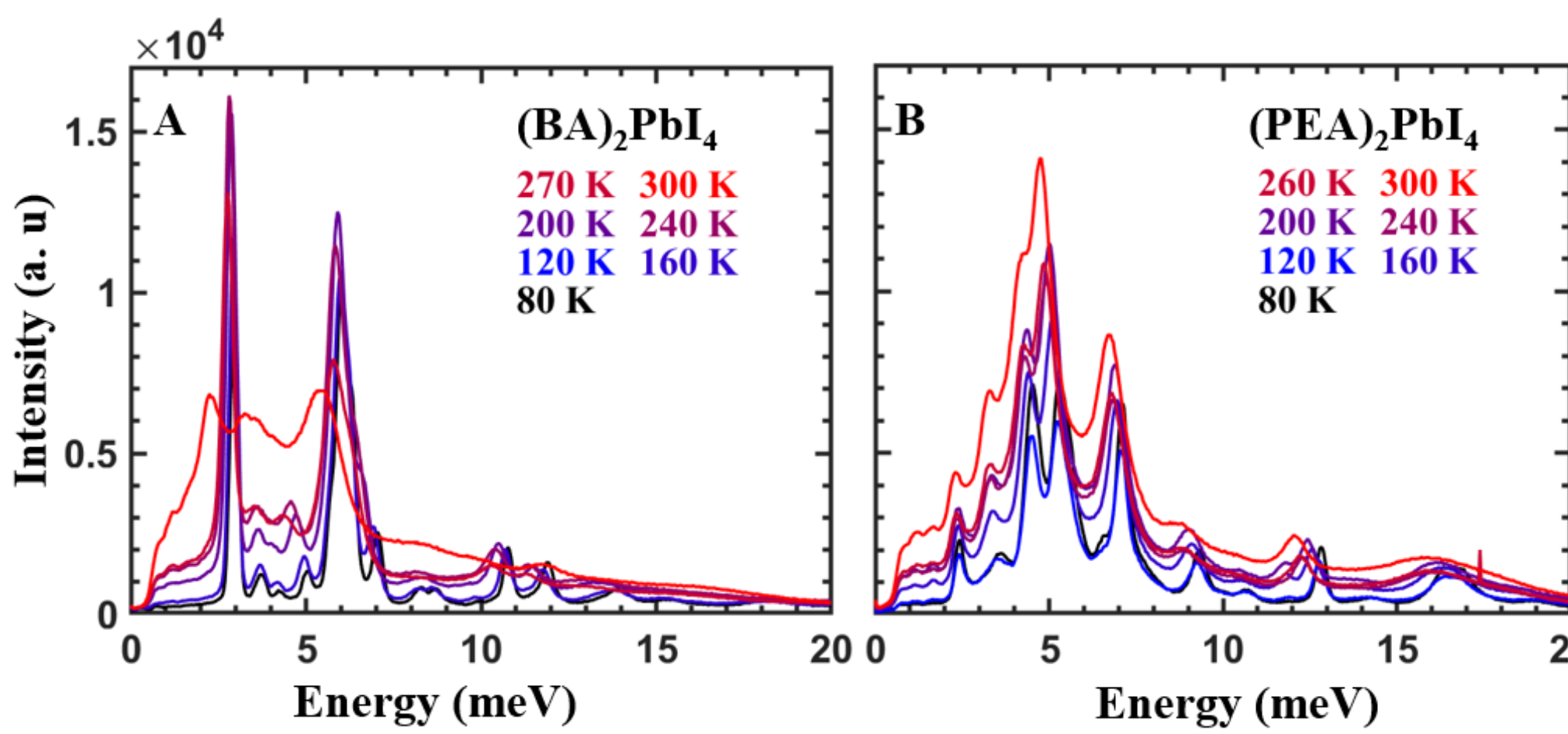

**FIG. 14** Temperature dependent Raman scattering spectra of $(BA)_2PbI_4$ (A) and $(PEA)_2PbI_4$ (B)

### 3.4. Photoluminescence

The photoluminescence (PL) spectra of powder samples of $(BA)_2PbI_4$ and $(PEA)_2PbI_4$ are shown in Figure 15 (A, B) and (E, F) respectively. At 100 K $(BA)_2PbI_4$ showed three emission peaks, one peak at ~2.4 eV ($P1_{BA}$) and two close peaks at ~ 2.53 eV ($P2_{BA}$) and ~2.55 eV ($P3_{BA}$) (Fig. 15 (C)). Literature studies intensively reported the presence of dual-excitonic emission peak in $(BA)_2PbI_4$ [57-60]. They predict that $(BA)_2PbI_4$ has a dual bandgap formed in the inside and the surface of the crystal - the surface emission ( $\sim 2.55$ eV ) and interior emission ( $\sim 2.4$ eV ). Our PL emissions confirm the presence of a dual band gap, except that the surface emission shows signatures of two peaks ($P2_{BA}$ and $P3_{BA}$) at 100 K (See Fig. S3 for specific temperature fitted curves). Up to 220 K the interior emission peak shifts to higher energies beyond which it starts shifting to lower energies. Both surface emission peaks shift to lower energies with an increase in temperature. We observed a sudden drop in both interior and surface emission peak energy above 275 K. This is expected due to the structural phase transition [50, 59].

$(PEA)_2PbI_4$ showed a single emission peak at low temperatures ($P1_{PEA}$) which then underwent a peak splitting above 200 K (Fig. 15(G)) The $P1_{PEA}$ shifts to lower emission energies with an increase in temperature. At 200 K a shoulder emission peak emerges at `~2.35 eV ($P2_{PEA}$), and the peak splitting becomes more prominent above 220 K (above a minimum energy separation of 0.03 eV) (See Fig. S4 for specific temperature fitted curves). Both peaks undergo thermal broadening upon warming (Fig. 15(H)). We emphasize that the emission peak splitting becoming prominent in both systems; T>150 K in the surface emission for $(BA)_2PbI_4$ and T>220 K for $(PEA)_2PbI_4$ occurs despite no change in crystal structure. On the other hand, it is interesting to note that the onset of decrease in PLQY upon heating occurs at similar temperatures for both systems. Additionally, our QENS data shows the activation of rotational dynamics at similar critical temperatures for both systems, which will be discussed in the following section. This suggests that molecular rotation might influence the emission and recombination mechanisms in 2D HOIPs. Notably, such PL peak splitting was not observed in the studies reported in the literature [61, 62]. We attribute this difference to the measurements being performed on single crystals or microplate crystals, whereas our measurements were done on powders.

Powder samples consist of many small crystallites with random orientations. This lack of long-range uniform structural order (compared to single crystals) and the presence of more surface and defect states introduce greater freedom for molecular rotations in the instrumental frame. This typically induces a variety of interactions between the material and incident light, leading to a broadening or even splitting of the PL spectra. This observation is consistent with the fact that we observe the splitting of the surface emission of $(BA)_2PbI_4$ powders but no splitting of the internal emission. In contrast, in single crystals, the uniformly ordered crystal structure results in more uniform interactions with incident light and fewer surface and defect states, typically leading to narrower peaks and better-defined PL spectra compared to powders. It has been demonstrated that different preparations of the same compound can strongly affect its optoelectronic properties; for instance, $(BA)_2PbI_4$ prepared using crystalline precursors and powder precursors shows dramatically different PL spectra [63]. Furthermore, in our work the PL peak splitting shows non-negligible temperature dependence, while pure effects from surface and defect states do not have such a significant T-dependence. Thus, it is highly likely that the T-dependent molecular rotations facilitate, not generate, the disturbances to PL spectra from surface and defect states. Powders have much more surface and defect states than single crystals, thus entanglements of PL spectra with molecular rotations become more pronounced.

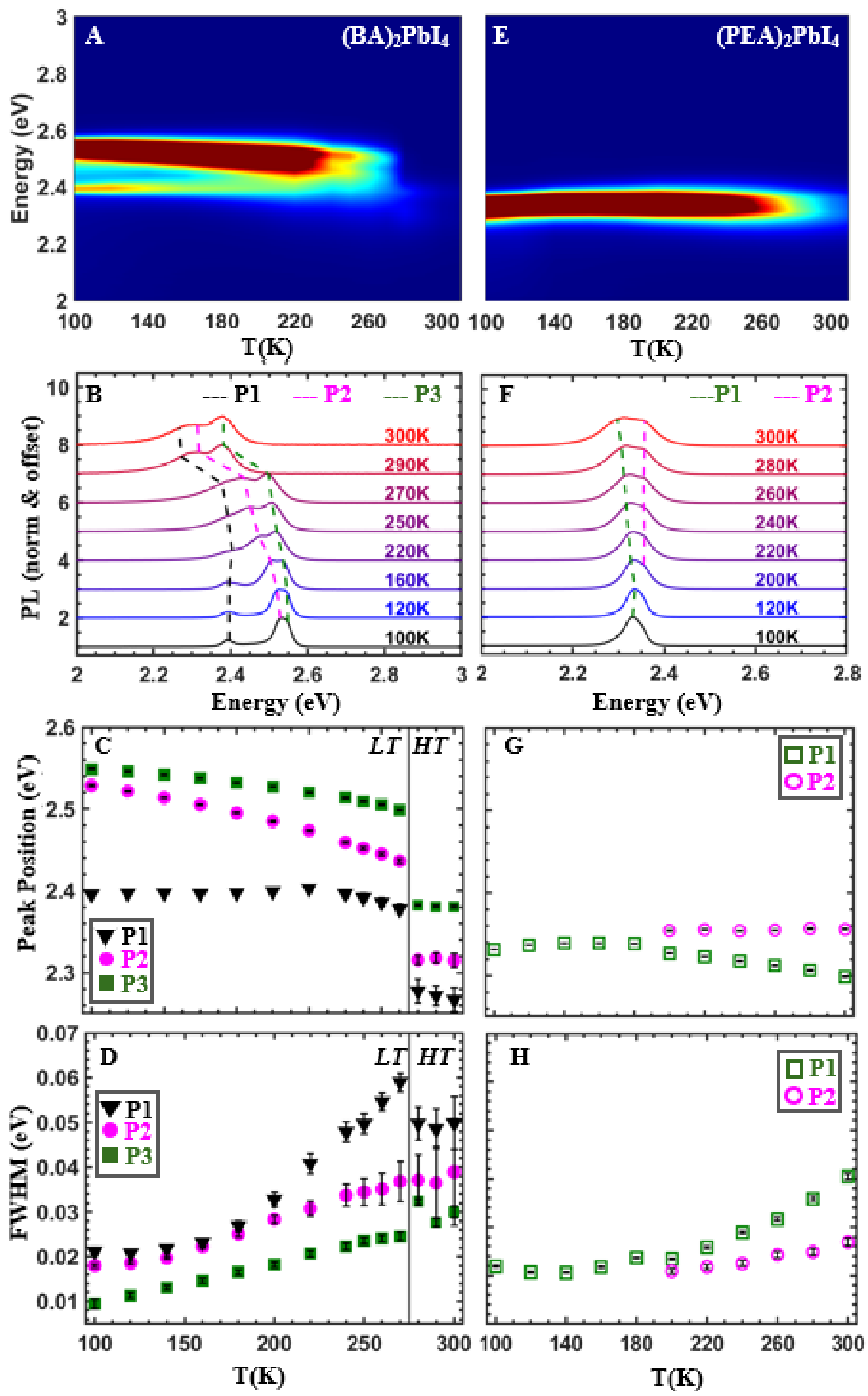

**FIG.15:** Temperature dependent PL Spectra for $(BA)_2PbI_4$( A)(B) and $(PEA)_2PbI_4$(E)(F). (C)(D) shows the PL emission peak position and peak width respectively for $(BA)_2PbI_4$ based on a triple peak fitting. (G)(H) shows the PL emission peak position and peak width respectively for $(PEA)_2PbI_4$ based on single peak fitting below 200K and dual peak fitting from 200K.

### 3.5. Discussions

In this work, to directly compare the temperature-dependent PLQY of $(BA)_2PbI_4$ and $(PEA)_2PbI_4$, we choose to evaluate the temperature-dependence of corresponding neutron scattering and Raman scattering phonon intensities, neutron scattering intensities from molecular rotations, and PL emission peak widths based on our previous analyses.

Figure 16 (A)(B) presents the temperature-dependence of inorganic and hybrid phonon intensities probed by Raman and neutron scattering. Both the Raman-probed and neutron-probed inorganic phonons ( $[1.5,4.0]$meV) show consistent temperature dependence - get enhanced monotonically upon warming, which validates our method of separating rotational and vibrational contributions in low-energy INS spectra (Fig. 16(A(I)), 16(B(I))). Interestingly the hybrid phonons ( $[12, 15]$meV) exhibit different behaviors in Raman and neutron scattering for $(BA)_2PbI_4$, while for $(PEA)_2PbI_4$ they are consistent (Fig. 16(A(II)), 16((B(II)). Comparing the temperature dependence of these phonon intensities with that of PLQY (Fig. 16(C)), it tends to suggest that the inorganic and hybrid phonons do not show explicit correlations with this characteristic opto-electronic property.

Figure 16 (C) showcases the rotational dynamics in these two systems in the energy range [1.5,4.0] meV, along with the PLQY of their bromide equivalents. Below 1.5 meV, there is little to no contribution from phonons, while above 4.0 meV, the contributions from rotations are negligible, even at the highest temperatures where rotations are most pronounced. By focusing on 1.5 to 4.0 meV, we aimed to assess a region where both phonons and rotations coexist and how it influences PLQY, providing a meaningful comparison of their contributions. Additionally, at low temperatures (below critical temperatures) for both $(BA)_2PbI_4$ and $(PEA)_2PbI_4$, the elastic-quasi-elastic region can be fully explained by the instrument resolution function (a Gaussian profile). However, above the critical temperature, the elastic channel peak broadens, which can be described by the Lorentzian component of the Voigt function. This broadening appears as a diffusion of intensity from the elastic channel into the quasi-elastic region, becoming more prominent beyond 1 meV away from the instrument resolution peak. This behavior is another reason why the energy region of 1.5 to 4.0 meV was chosen for analysis.

Below $\sim 140$ K and $\sim 220$ K, respectively, the molecular rotations in $(BA)_2PbI_4$ and $(PEA)_2PbI_4$ are 'frozen', manifested as negligible QENS intensities. Upon heating above these critical temperatures up to 300 K, the rotational dynamics of BA molecules gets greatly enhanced while that of PEA molecules gets slightly strengthened, which is inversely proportional to the temperature dependence of PLQY for both materials. These observations indicate that the molecular rotations in these two HOIPs have non-negligible influences on the intrinsic opto-electronic property. Additionally, the signatures of PL emission peak splitting becoming more prominent above 150 K for $(BA)_2PbI_4$ and 220 K for $(PEA)_2PbI_4$ can also be observed as a more dynamic broadening of the emission peaks $P2_{BA}$ and $P2_{PEA}$ (Fig. 16(D)) above these critical temperatures. Both $P2_{BA}$ and $P2_{PEA}$ become shoulder peaks to the emission peaks $P3_{BA}$ and $P1_{PEA}$ respectively above their corresponding critical temperatures. The peak splitting could be an indication of breaking of the degeneracy of exciton binding energy levels. This could be due to some local perturbation of the dielectric environment around the exciton. Such changes in the dielectric environment could also affect the exciton binding energy. But given the complex dynamics in 2D HOIPs and the discrepancies in the reported and calculated exciton binding energies in these systems [64], quantifying them will not be of interest of this paper. Thus, combined with the observations on lattice vibrations, we could propose that: The rotational motion of polarized organic molecules works as dynamical perturbations to the inorganic framework which fundamentally dominates the construction of electronic band structure. These perturbations interfere with the dielectric environment surrounding excitons, potentially reducing the exciton binding energy or breaking the binding energy degeneracy, which can enhance the non-radiative decay of charge carriers and, consequently, suppress the photoluminescence quantum yield (PLQY). To confirm this hypothesis temperature dependent dielectric measurements can be done for these two systems as a next step. More detailed investigations with single-crystals on molecular orientations could reveal the detailed mechanism underlying this rotation-induced optoelectronic effect.

## 4. Conclusions

In summary, our work studied the influence of structural dynamics in two 2D HOIPs, $(BA)_2PbI_4$ and $(PEA)_2PbI_4$, on their photoluminescence efficiency. The temperature dependent X-ray diffraction analysis indicates that the distortion of inorganic layers is mainly associated with structural phase transition and does not correlate explicitly with their PLQY. By using QENS and group theory we quantitatively identified the rotational modes of different rotors in these two HOIPs. We carefully separated the rotational intensity contributions and vibrational contributions in the total INS spectra. Complemented by Raman scattering measurements on lattice vibrations and photoluminescence measurements on emission characteristics, we argued that in these two 2D HOIPs the molecular rotations have non-negligible impacts on their intrinsic opto-electronic property (PLQY) while lattice vibrations do not show explicit correlations. These findings could make essential contributions to the big picture of 2D HOIPs optoelectronics and may guide the design of more efficient light-emitting materials for advanced technological applications.

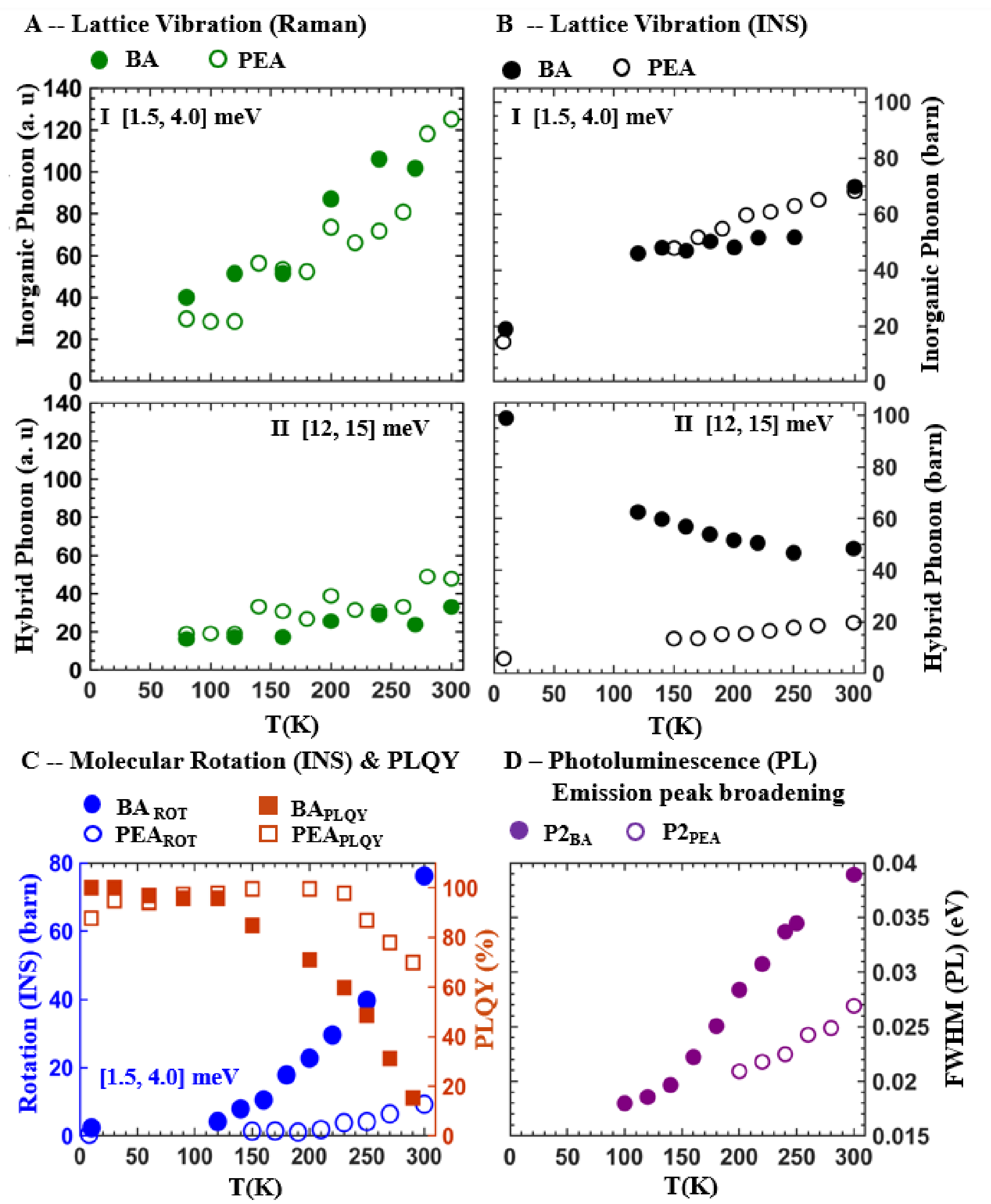

**FIG. 16**: (A) Temperature-dependent integrated phonon intensities from Raman scattering for $(BA)_2PbI_4$ and $(PEA)_2PbI_4$ over the energy range of $1.5 \leq \hbar\omega \leq 4.0\text{meV}$ (I) and $12 \leq \hbar\omega \leq 15\text{meV}$ (II) and a Q range of $1 \sim 2.5\text{Å}^{-1}$. (B) Temperature-dependent integrated phonon intensities from INS for $(BA)_2PbI_4$ and $(PEA)_2PbI_4$ over the energy range of $1.5 \leq \hbar\omega \leq 4.0\text{meV}$ (I) and $12 \leq \hbar\omega \leq 15\text{meV}$ (II) and the same Q range. (C) Blue circles show the rotational intensity contributions of $(BA)_2PbI_4$ and $(PEA)_2PbI_4$ from INS spectra over the energy range of $1.5 \leq \hbar\omega \leq 4.0\text{meV}$ and same Q range. Orange squares represent the temperature dependent PLQY of $(BA)_2PbBr_4$ and $(PEA)_2PbBr_4$ reported by Ref. [33]. (D) Temperature dependent broadening PL emission peaks for $(BA)_2PbI_4$ ($P2_{BA}$) and $(PEA)_2PbI_4$ ($P2_{PEA}$).

## Acknowledgements

The work at University of Virginia (H. S. R., X. H., S-H. L.) was supported by the US Department of Energy, Office of Science, Office of Basic Energy Sciences under the Award No: DE-SC001644. The neutron scattering experiments at the Material Science and Life Science Experimental facility, Japan Proton Accelerator Research Complex, were performed under the user programs with proposal no: 2017AU0101 and 2018B02028 for 4SEASONS and 2017BU1402 and 2018B0257 for AMATERAS (R. K., Ma. Ko., M. N., D. Z., T. C.). Raman scattering and photoluminescence measurements were performed at the Center for Condensed Matter Sciences, National Taiwan University (W-L. C, Y-M. C.). Theory work was partly supported by the U.S. Department of Energy, Office of Science, Office of Basic Energy Sciences, Materials Sciences and Engineering Division (M. Y.). Computational work used resources of the Oak Ridge Leadership Computing Facility at the Oak Ridge National Laboratory, which is supported by the Office of Science of the U.S. Department of Energy under Contract No. DE-AC05-00OR22725 and resources of the National Energy Research Scientific Computing Center, a DOE Office of Science User Facility supported by the Office of Science of the U.S. Department of Energy under Contract No. DE-AC02-05CH11231 using NERSC award BES-ERCAP0024568. Samples were prepared by A.Z.C. and G.C.J.

## References

1. A. Mei, X. Li, L. Liu, Z. Ku, T. Liu, Y. Rong, M. Xu, M. Hu, J. Chen, Y. Yang, M. Grätzel, and H. Han, A hole-conductor-free, fully printable mesoscopic perovskite solar cell with high stability. *Science* **345**(6194), 295–298 (2014)
2. E. H. Jung, N. J. Jeon, E. Y. Park, C. S. Moon, T. J. Shin, T.-Y. Yang, J. H. Noh, and J. Seo, Efficient, stable and scalable perovskite solar cells using poly (3-hexylthiophene). *Nature* **567**, 511 (2019)
3. W. S. Yang, B.-W. Park, E. H. Jung, N. J. Jeon, Y. C. Kim, D. U. Lee, S. S. Shin, J. Seo, E. K. Kim, J. H. Noh, and S. I. Seok, Iodide management in formamidinium lead-halide-based perovskite layers for efficient solar cells. *Science* **356**, 1376 (2017)
4. Q. Jiang, Y. Zhao, X. Zhang, X. Yang, Y. Chen, Z. Chu, Q. Ye, X. Li, Z. Yin, and J. You, Surface passivation of perovskite film for efficient solar cells. *Nat. Photonics*, **13**(7), 460–466 (2019)
5. S. Yang, S. Chen, E. Mosconi, Y. Fang, X. Xiao, C. Wang, Y. Zhou, Z. Yu, J. Zhao, Y. Gao, F. De Angelis, and J. Huang, Stabilizing halide perovskite surfaces for solar cell operation with wide-bandgap lead oxysalts. *Science* **365**(6452), 473–478 (2019)
6. T. M. Brenner, D. A. Egger, L. Kronik, G. Hodes, and D. Cahen, Hybrid organic-inorganic perovskites: Low-cost semiconductors with intriguing charge-transport properties, *Nat. Rev. Mater.* **1**, 15007 (2016)
7. W. Xu, Q. Hu, S. Bai, C. Bao, Y. Miao, Z. Yuan, T. Borzda, A. J. Barker, E. Tyukalova, Z. Hu, M. Kawecki, H. Wang, Z. Yan, X. Liu, X. Shi, K. Uvdal, M. Fahlman, W. Zhang, M. Duchamp, J.-M. Liu, A. Petrozza, J. Wang, L.-M. Liu, W. Huang, and F. Gao, Rational molecular passivation for high-performance perovskite light-emitting diodes. *Nat. Photonics* **13**, 418 (2019)

8. L. N. Quan, F. P. Garcia de Arquer, R. P. Sabatini, and E. H. Sargent, Perovskites for light emission, *Adv. Mater.* **30**(45), 1801996 (2018)
9. Z.-K. Tan, R. S. Moghaddam, M. L. Lai, P. Docampo, R. Higler, F. Deschler, M. Price, A. Sadhanala, L. M. Pazos, D. Credgington, F. Hanusch, T. Bein, H. J. Snaith, and R. H. Friend, Bright light-emitting diodes based on organometal halide perovskite, *Nat. Nanotechnol.* **9**(9), 687–692 (2014)
10. G. Xing, N. Mathews, S. S. Lim, N. Yantara, X. Liu, D. Sabba, M. Grätzel, S. Mhaisalkar, and T. C. Sum, Low-temperature solution-processed wavelength-tunable perovskites for lasing, *Nat. Mater.* **13**(5), 476–480 (2014)
11. H. Zhu, Y. Fu, F. Meng, X. Wu, Z. Gong, Q. Ding, M. V. Gustafsson, M. T. Trinh, S. Jin, and X.-Y. Zhu, Lead halide perovskite nanowire lasers with low lasing thresholds and high-quality factors, *Nat. Mater.* **14**(6), 636–642 (2015)
12. C. Li, H. Wang, F. Wang, T. Li, M. Xu, H. Wang, Z. Wang, X. Zhan, W. Hu, and L. Shen, Ultrafast and broadband photodetectors based on a perovskite/organic bulk heterojunction for large-dynamic-range imaging, *Light Sci Appl* **9**, 31 (2020)
13. H. Jing, R. Peng, R.-M. Ma, J. He, Y. Zhou, Z. Yang, C.-Y. Li, Y. Liu, X. Guo, Y. Zhu, D. Wang, J. Su, C. Sun, W. Bao, and M. Wang, Flexible ultrathin single-crystalline perovskite photodetector, *Nano Letters* **20**(10), 7144-7151 (2020)
14. K. Miyata, T. L. Atallah, and X. Y. Zhu, Lead halide perovskites: Crystal-liquid duality, phonon glass electron crystals, and large polaron formation, *Sci. Adv.* **3**, e1701469 (2017)
15. G. Xing, N. Mathews, S. Sun, S. S. Lim, Y. M. Lam, M. Grätzel, S. Mhaisalkar, and T. C. Sum, Long-range balanced electron and hole-transport lengths in organic-inorganic $CH_3NH_3PbI_3$, *Science* **342**, 344 (2013)
16. Q. Dong, Y. Fang, Y. Shao, P. Mulligan, J. Qiu, L. Cao, and J. Huang, Electron-hole diffusion lengths >175 μm in solution-grown $CH_3NH_3PbI_3$ single crystals, *Science* **347**, 967 (2015)
17. D. Shi, V. Adinolfi, R. Comin, M. Yuan, E. Alarousu, A. Buin, Y. Chen, S. Hoogland, A. Rothenberger, K. Katsiev, Y. Losovyj, X. Zhang, P. A. Dowben, O. F. Mohammed, E. H. Sargent, and O. M. Bakr, Low trap-state density and long carrier diffusion in organolead trihalide perovskite single crystals, *Science* **347**, 519 (2015)
18. D. Emin, "*Polarons*", Cambridge: Cambridge University Press (2012)
19. T. Chen, W.-L. Chen, B. J. Foley, J. Lee, J. P. C. Ruff, J. Y. Peter Ko, C. M. Brown, L. W. Harriger, D. Zhang, C. Park, M. Yoon, Y.-M. Chang, J. J. Choi, and S.-H. Lee, Origin of long lifetime of band-edge charge carriers in organic-inorganic lead iodide perovskites, *PNAS* **114**, 7519 (2017)
20. K. Wang, Z. Jin, L. Liang, H. Bian, D. Bai, H. Wang, J. Zhang, Q. Wang, and S. Liu, All-inorganic cesium lead iodide perovskite solar cells with stabilized efficiency beyond 15%, *Nat. Commun.* **9**, 4544 (2018)
21. K. Wang, Z. Jin, L. Liang, H. Bian, H. Wang, J. Feng, Q. Wang, and S. Liu, Chlorine doping

for black $\gamma$- $CsPbI_3$ solar cells with stabilized efficiency beyond 16%, *Nano Energy* **58**, 175 (2019)
22. K. Miyata, D. Meggiolaro, M. T. Trinh, P. P. Joshi, E. Mosconi, S. C. Jones, F. De Angelis, and X. Y. Zhu, Large polarons in lead halide perovskites, *Sci. Adv.* **3**, e1701217 (2017)
23. H. H. Fang, L. Protesescu, D. M. Balazs, S. Adjokatse, M. V. Kovalenko, and M. A. Loi, Exciton recombination in formamidinium lead triiodide: nanocrystals versus thin films, *Small* **13**, 1700673 (2017)
24. S. Tombe, G. Adam, H. Heilbrunner, D. H. Apaydin, C. Ulbricht, N. S. Sariciftci, C. J. Arendse, E. Iwuoha, and M. C. Scharber, Optical and electronic properties of mixed halide (X = I, Cl, Br) methylammonium lead perovskite solar cells, *J. Mater. Chem. C* **5**, 1714 (2017)
25. H. Kim, J. Hunger, E. Cánovas, M. Karakus, Z. Mics, M. Grechko, D. Turchinovich, S. H. Parekh, and M. Bonn, Direct observation of mode-specific phonon-band gap coupling in methylammonium lead halide perovskites, *Nat. Commun.* **8**, 687 (2017)
26. S. Zhang, C. Yi, N. Wang, Y. Sun, W. Zou, Y. Wei, Y. Cao, Y. Miao, R. Li, Y. Yin, N. Zhao, J. Wang, and W. Huang, Efficient red perovskite light-emitting diodes based on solution-processed multiple quantum wells, *Adv. Mater.* **29**, 1606600 (2017)
27. B. Traore, L. Pedesseau, L. Assam, X. Che, J.-C. Blancon, H. Tsai, W. Nie, C. C. Stoumpos, M. G. Kanatzidis, S. Tretiak, A. D. Mohite, J. Even, M. Kepenekian, and C. Katan, Composite nature of layered hybrid perovskites: assessment on quantum and dielectric confinements and band alignment. *ACS Nano* **12**, 3321-3332 (2018)
28. C. Katan, N. Mercier, and J. Even, Quantum and dielectric confinement effects in lower-dimensional hybrid perovskite semiconductors. *Chem. Rev.* **119**, 3140-3192 (2019)
29. N. Wang, L. Cheng, R. Ge, S. Zhang, Y. Miao, W. Zou, C. Yi, Y. Sun, Y. Cao, R. Yang, Y. Wei, Q. Guo, Y. Ke, M. Yu, Y. Jin, Y. Liu, Q. Ding, D. Di, L. Yang, G. Xing, H. Tian, C. Jin, F. Gao, R. H. Friend, J. Wang, and W. H, Perovskite light-emitting diodes based on solution-processed self-organized multiple quantum wells. *Nat. Photonics* **10**, 699-704 (2016)
30. Z. He, Y. Liu, Z. Yang, J. Li, J. Cui, D. Chen, Z. Fang, H. He, Z. Ye, H. Zhu, N. Wang, J. Wang, and Y. Jin, High-efficiency red light-emitting diodes based on multiple quantum wells of phenylbutylammonium-cesium lead iodide perovskites. *ACS Photonics* **6**, 587-594 (2019)
31. J.-C. Blancon, A. V. Stier, H. Tsai, W. Nie, C. C. Stoumpos, B. Traoré, L. Pedesseau, M. Kepenekian, F. Katsutani, G. T. Noe, J. Kono, S. Tretiak, S. A. Crooker, C. Katan, M. G. Kanatzidis, J. J. Crochet, J. Even, and A. D. Mohite, Scaling law for excitons in 2D perovskite quantum wells, *Nat. Commun.* **9**, 2254 (2018)
32. E. Da Como, F. De Angelis, H. Snaith, and A. Walker, Unconventional Thin Film Photovoltaics, Royal Society of Chemistry, (2016).
33. X. Gong, O. Voznyy, A. Jain, W. Liu, R. Sabatini, Z. Piontkowski, G. Walters, G. Bappi, S. Nokhrin, O. Bushuyev, M. Yuan, R. Comin, D. McCamant, S. O. Kelley, E. H. Sargent, Electron-phonon interaction in efficient perovskite blue emitters, *Nat. Mater.* **17**, 550-556 (2018).

34. M. Bée, Quasielastic Neutron Scattering: Principles and Applications in Solid State Chemistry, Biology and Materials Science, *Bristol and Philadelphia: Adam Hilger* (1988)
35. T. Chen, B. J. Foley, B. Ipek, M. Tyagi, J. R. D. Copley, C. M. Brown, J. J. Choi, and S.-H. Lee, Rotational dynamics of organic cations in $CH_3NH_3PbI_3$ perovskite, *Phys. Chem. Chem. Phys.* **17**, 31278 (2015)
36. Bruker, APEX SAINT, and Bruker AXS Inc. "Madison, Wisconsin, USA, 2004, (b) Sheldrick." *Acta Crystallographica Section A*, 46, 467-473 (1990).
37. K. Nakajima, S. O. Kawamura, T. Kikuchi, M. Nakamura, R. Kajimoto, Y. Inamura, N. Takahashi, K. Aizawa, K. Suzuya, K. Shibata, T. Nakatani, K. Soyama, R. Maruyama, H. Tanaka, W. Kambara, T. Iwahashi, Y. Itoh, T. Osakabe, S. Wakimoto, K. Kakurai, F. Maekawa, M. Harada, K. Oikawa, R. E. Lechner, F. Mezei, and M. Arai, AMATERAS: A cold-neutron disk chopper spectrometer, *Journal of the Physical Society of Japan* **80**(Suppl. B): p. SB028 (2011)
38. R. Kajimoto, M. Nakamura, Y. Inamura, F. Mizuno, K. Nakajima, S. O. Kawamura, T. Yokoo, T. Nakatani, R. Maruyama, K. Soyama, K. Shibata, K. Suzuya, S. Sato, K. Aizawa, M. Arai, S. Wakimoto, M. Ishikado, S. Shamoto, M. Fujita, H. Hiraka, K. Ohoyama, K. Yamada, and C.-H. Lee, The Fermi chopper spectrometer 4SEASONS at J-PARC, *Journal of the Physical Society of Japan* **80**(Suppl. B): p. SB025 (2011)
39. Nakamura, Mitsutaka, Ryoichi Kajimoto, Yasuhiro Inamura, Fumio Mizuno, Masaki Fujita, Tetsuya Yokoo, and Masatoshi Arai, First demonstration of novel method for inelastic neutron scattering measurement utilizing multiple incident energies, Journal *of the Physical Society of Japan* **78**, p.093002 (2009).
40. G. Kresse, and J. Furthmüller. Efficiency of ab-initio total energy calculations for metals and semiconductors using a plane-wave basis set. *Computational materials science*, **6**, 15-50 (1996)
41. Kresse, G., and Joubert D. From ultrasoft pseudopotentials to the projector augmented-wave method. *Physical review B*, **59**, 1758-1775 (1999)
42. Perdew, J.P., Burke, K., and Ernzerhof, M. Generalized gradient approximation made simple. *Phys. Rev. Lett*, **77**, 3865-3868 (1996)
43. Momma, K. and Izumi, F., VESTA 3 for three-dimensional visualization of crystal, volumetric and morphology data. *Journal of Applied Crystallography*, **44**, 1272-1276 (2011)
44. Billing, D.G. and Lemmerer, A. Synthesis, characterization and phase transitions in the inorganic–organic layered perovskite-type hybrids $[(C_n\mathrm{H}_{2n+1}\mathrm{NH}_3)_2\mathrm{PbI}_4]$, n= 4, 5 and 6. *Acta Crystallographica Section B: Structural Science*, **63**, 735-747 (2007)
45. Xiao, Y., Xue, C., Wang, X., Liu, Y., Yang, Z. and Liu, S. Bulk Heterostructure $BA_2PbI_4$/ $MAPbI_3$ Perovskites for Suppressed Ion Migration to Achieve Sensitive X-ray Detection Performance. ACS Applied Materials & Interfaces, 14(49), 54867-54875 (2022)
46. Cao, Y., Wang, Y., Zhang, Y., Liu, X. and Lin, J. Stable and rapid thermochromic reversibility in acene alkylamine intercalated layered hybrid perovskites. Materials Today Communications, 37, 107544 (2023)

47. Du, K.Z., Tu, Q., Zhang, X., Han, Q., Liu, J., Zauscher, S. and Mitzi, D.B. Two-dimensional lead (II) halide-based hybrid perovskites templated by acene alkylamines: crystal structures, optical properties, and piezoelectricity. *Inorganic chemistry*, **56**, 9291-9302 (2017)
48. Kowal, D., Makowski, M., Witkowski, M.E., Calà, R., Sheikh, M.K., Mahyuddin, M.H., Auffray, E., Drozdowski, W., Cortecchia, D. and Birowosuto, M.D. $(PEA)_2PbI_4$: fast two-dimensional lead iodide perovskite scintillator with green and red emission. Materials Today Chemistry, 29, 101455 (2023)
49. Bonomi, S., Armenise, V., Accorsi, G., Colella, S., Rizzo, A., Fracassi, F., Malavasi, L. and Listorti, A. The effect of extended ball-milling upon three-dimensional and two-dimensional perovskite crystals properties. Applied Sciences, 10(14), 4775 (2020)
50. J. D. Ziegler, K. Q. Lin, B. Meisinger, X. Zhu, M. Kober-Czerny, P. K. Nayak, ... & A. Chernikov. Excitons at the phase transition of 2D hybrid perovskites. *ACS Photonics*, **9**, 3609-3616 (2022).
51. M. Dyksik, H. Duim, X. Zhu, Z. Yang, M. Gen, Y. Kohama, ... & P. Plochocka. Broad tunability of carrier effective masses in two-dimensional halide perovskites. ACS Energy Letters, 5(11), 3609-3616 (2020).
52. Toby, B.H., and Von Dreele, R.B. "GSAS-II: The genesis of a modern open-source all-purpose crystallography software package." *Journal of Applied Crystallography*, 46(2), 544-549 (2013).
53. Chen, T., Foley, B.J., Park, C., Brown, C.M., Harriger, L.W., Lee, J., Ruff, J., Yoon, M., Choi, J.J., and Lee, S.H. Entropy-driven structural transition and kinetic trapping in formamidinium lead iodide perovskite. *Sci. Adv.* **2**, e1601650 (2016).
54. Y. Q. Cheng, L. L. Daemen, A. I. Kolesnikov, and A. J. Ramirez-Cuesta, Simulation of inelastic neutron scattering spectra using OCLIMAX. *J. Chem. Theory Comput.* **15**, 3, 1974-1982 (2019).
55. G. Xu, Z. Xu, and J. M. Tranquada, Absolute cross-section normalization of magnetic neutron scattering data, *Review of Scientific Instruments* **84**, 083906 (2013)
56. Anderson, Anthony. "The Raman Effect. Volume 1: Principles." (1972)
57. Sheikh, T., Nawale, V., Pathoor, N., Phadnis, C., Chowdhury, A., and Nag, A. Molecular intercalation and electronic two dimensionality in layered hybrid perovskites. *Angew. Chem.* 132, 11750-11756 (2020).
58. DeCrescent, R.A., Du, X., Kennard, R.M., Venkatesan, N.R., Dahlman, C.J., Chabinyc, M.L., and Schuller, J.A. Even-parity self-trapped excitons lead to magnetic dipole radiation in two-dimensional lead halide perovskites. *ACS Nano* 14, 8958-8968 (2020).
59. Wang, Y., He, C., Tan, Q., Tang, Z., Huang, L., Liu, L., Yin, J., Jiang, Y., Wang, X., and Pan, A. Exciton–phonon coupling in two-dimensional layered $(BA)_2PbI_4$ perovskite microplates. *RSC Adv.* **13**, 5893-5899 (2023).
60. Du, Q., Zhu, C., Yin, Z., Na, G., Cheng, C., Han, Y., Liu, N., Niu, X., Zhou, H., Chen, H., and Zhang, L. Stacking effects on electron–phonon coupling in layered hybrid perovskites via microstrain manipulation. *ACS Nano* 14, 5806-581720).

61. Hong, H., Guo, S., Jin, L., Mao, Y., Chen, Y., Gu, J., Chen, S. et al. Two-dimensional lead halide perovskite lateral homojunctions enabled by phase pinning. *Nature Communications*, **15**(1), 3164 (2024).
62. Wang, J., Shen, H., Li, J., Li, W., Fang, C., Ma, J., Cheng, X., and Li, D. Electric-field-induced phase transition in 2D layered perovskite $(BA)_2PbI_4$ microplate crystals. *Applied Physics Letters*, **116**(5) (2020).
63. Gélvez-Rueda, M.C., Peeters, S., Wang, P.-C., Felter, K.M., and Grozema, F.C. Effect of structural defects and impurities on the excited state dynamics of 2D $BA_2PbI_4$ perovskite. Helv. Chim. Acta 103, e2000121 (2020).
64. Forde, A., Tretiak, S. and Neukirch, A.J. Dielectric Screening and Charge-Transfer in 2D Lead-Halide Perovskites for Reduced Exciton Binding Energies. Nano Letters, 23(24), 11586-11592 (2023)

# SUPPLEMENTARY INFORMATION

**The influence of Structural Dynamics in Two-Dimensional Hybrid Organic-Inorganic Perovskites on their Photoluminescence Efficiency – Neutron scattering analysis**

Haritha Sindhu Rajeev, [1,*] Xiao Hu, [1,†] Wei-Liang Chen, [2] Depei Zhang, [1,‡] Tianran Chen, [1,||] Maiko Kofu, [3] Ryoichi Kajimoto, [3] Mitsutaka Nakamura, [3] Alexander Z. Chen, [4] Grayson C. Johnson, [4] Mina Yoon, [5] Yu-Ming Chang, [2] Diane A. Dickie, [6] Joshua J. Choi, [4] and Seung-Hun Lee [1]

[1] *Department of Physics, University of Virginia, Charlottesville, Virginia 22904, USA*

† *Present address: Condensed Matter Physics and Materials Science Division, Brookhaven National Laboratory, Upton, NY 11973, USA*

[2] *Center for Condensed Matter Sciences, National Taiwan University, Taipei 10617, Taiwan*

‡ *Present address: Currently working in Robinhood, USA*

|| *Present address: Department of Physics and Astronomy, University of Tennessee, Knoxville TN 37996, USA*

[3] *J-PARC Center, Japan Atomic Energy Agency, Ibaraki 319-1195, Japan*

[4] *Department of Chemical Engineering, University of Virginia, Charlottesville, Virginia 22903, USA*

[5] *Center for Nanophase Materials Sciences, Oak Ridge National Laboratory, Oak Ridge TN 37831, USA*

[6] *Department of Chemistry, University of Virginia, Charlottesville, Virginia 22904, USA*

Corresponding author. hs8bp@virginia.edu

**This file includes:**

1. XRD refinement details
2. Vibrational dynamics details
3. Photoluminescence spectral fitting details

Figs. S1 to S5

## 1. XRD Refinement details

The refinement of XRD patterns was performed using GSAS-II. To quantify the octahedral tilts, the refinement employed the restraint method, allowing bond lengths and bond angles within the inorganic octahedral layer to vary within a specified range with carefully chosen weight factors. These weight factors are critical for balancing the influence of restraints and ensuring the refined structure remains physically realistic while fitting the experimental data. To ensure the model fits well to the experimental data, we used Rw (weighted R-factor) values as a measure of fit quality. Additionally, to avoid overfitting due to the number of refined parameters, an Rw value below 10% was considered ideal for a reliable fit. Properly adjusted weight factors also help prevent overfitting, which could otherwise result in unrealistically low Rw values and refined structures with bond lengths and angles that deviate significantly from realistic values. The refined lattice parameters with corresponding errors are presented in Table I of the main text. Only the patterns at the highest (300 K) and lowest (80 K) temperatures are shown here, as the intermediate temperatures yielded similar results. For $(BA)_2PbI_4$, the patterns remain consistent in the low-temperature (LT) phase ($T<275$ K), and for $(PEA)_2PbI_4$, no significant changes are observed across the entire temperature range (80 K to 300 K). This consistency is evident from the octahedral out-of-plane tilt angles ($\delta$) reported in Fig. 3(B) of the main text, where minimal variation in $\delta$ is observed for $(PEA)_2PbI_4$) and the LT phase of $(BA)_2PbI_4$.

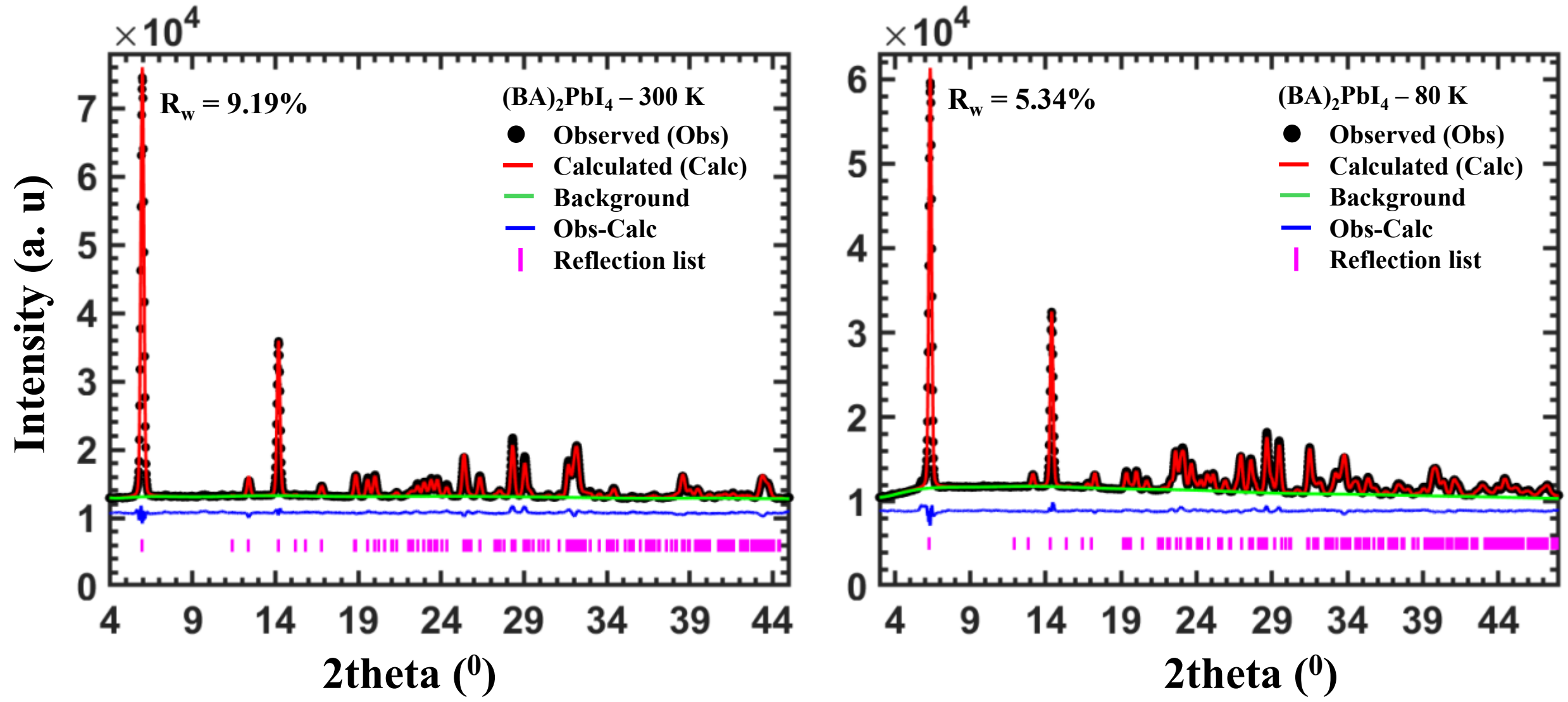

**FIG S1:** "Color" XRD refined patterns for $(BA)_2PbI_4$ obtained using GSAS-II at 300 K and 80 K.

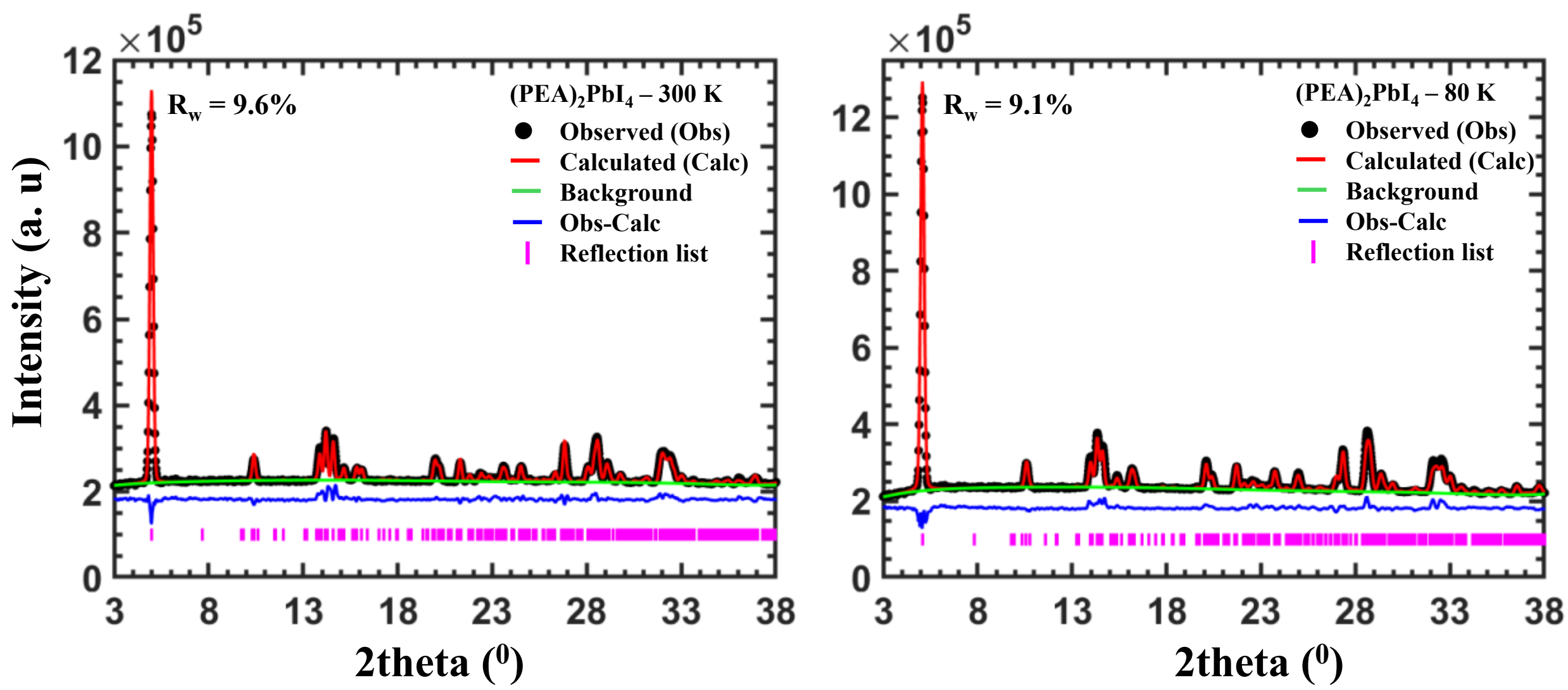

**FIG S2:** "Color" XRD refined patterns for $(PEA)_2PbI_4$ obtained using GSAS-II at 300 K and 80 K.

## 2. Vibrational dynamics details

Based on the DFT calculations, we characterized the phonon modes into three different types: inorganic phonon modes, hybrid phonon modes, and organic phonon modes in terms of the vibration energy fraction from different atoms in the following way. The vibrational energy fraction $V_{AT}(s,\ q)$ is defined as the energy contribution of each atomic type (***AT***) within the unit cell, encompassing the atoms Pb, I, N, C, and H.

$$V_{AT}(s,\ q) = \frac{\sum_{i=1}^{n_{AT}} m_i \omega_s(q)^2 |u_i(s,q)|^2}{\sum_{AT\ \epsilon\ atoms} \sum_{i=1}^{n_{AT}} m_i \omega_s(q)^2 |u_i(s,q)|^2}$$

(Eq. S1)

Where $s$ and $q$ represent the phonon mode index and phonon wavevector respectively. $n_{AT}$ is the number of a specific atomic type in the unit cell, $|u_i(s,q)|^2$ is the mean square displacement of the i-th atom due to the activation of phonon mode $s$. $\hbar$ is the reduced Planck constant, $m_i$ is the mass of the i-th atom and $\omega_s(q)$ is the eigen frequency of phonon mode $s$ at $q$.

The vibrational energy fractions at Γ point of the inorganic layer, $V_{PbI}$ were calculated by summing up the vibrational energy fraction of Pb ($V_{Pb}$ ) and vibrational energy fraction of I ($V_I$ ). Fig S1 shows the histogram of number of phonon modes with certain inorganic Pb-I energy fractions $[V_{PbI}\ ,\ V_{PbI} + \Delta V]$ for both systems, where $\Delta V$ represents the histogram bin width, which is 2%. The modes are categorized into inorganic phonons modes $V_{PbI}$ > 70% for $(BA)_2PbI_4$ and $V_{PbI}$> 65% for $(PEA)_2PbI_4$, organic phonons when $V_{PbI}$ < 2%, and hybrid phonons in between. The vibrational energy fractions of other atomic types C, H, and N were also calculated, and the contributions from all atomic types to each phonon mode at the Γ point for $(BA)_2PbI_4$ and $(PEA)_2PbI_4$ are shown in **Figs. 9, 10 panels (D) of the main text**. Most of the inorganic phonons lie in the energy range of $\hbar\omega$ < 10 meV; the hybrid phonons mostly locate in the energy range of 10~35 meV; and the energies of organic phonons usually range from a few tens to hundreds of meV.

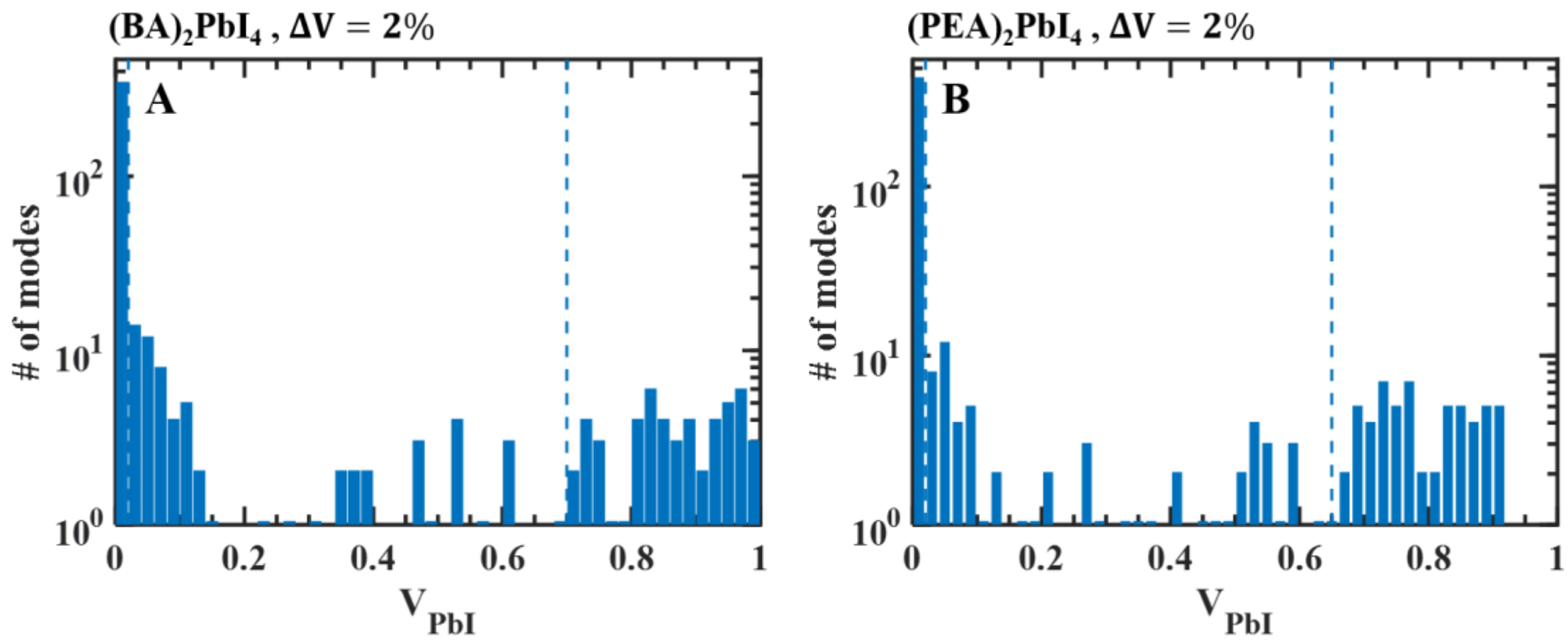

**FIG. S3** "Color" Statistics of number of phonon modes with different Pb-I vibrational energy fractions for $(BA)_2PbI_4$ and $(PEA)_2PbI_4$. The vertical bars represent the number of phonon modes with certain Pb-I energy fractions $[V_{PbI}, V_{PbI} + \Delta V]$. The minimum non-zero value is 1. Vertical dashed lines represent the threshold Pb-I energy fractions that separate the inorganic modes ($V_{PbI} > 70\%$ for $(BA)_2PbI_4$ and 65% for $(PEA)_2PbI_4$) from organic modes ($V_{PbI} < 2\%$), and hybrid modes in between.

## 3. Photoluminescence spectral fitting details

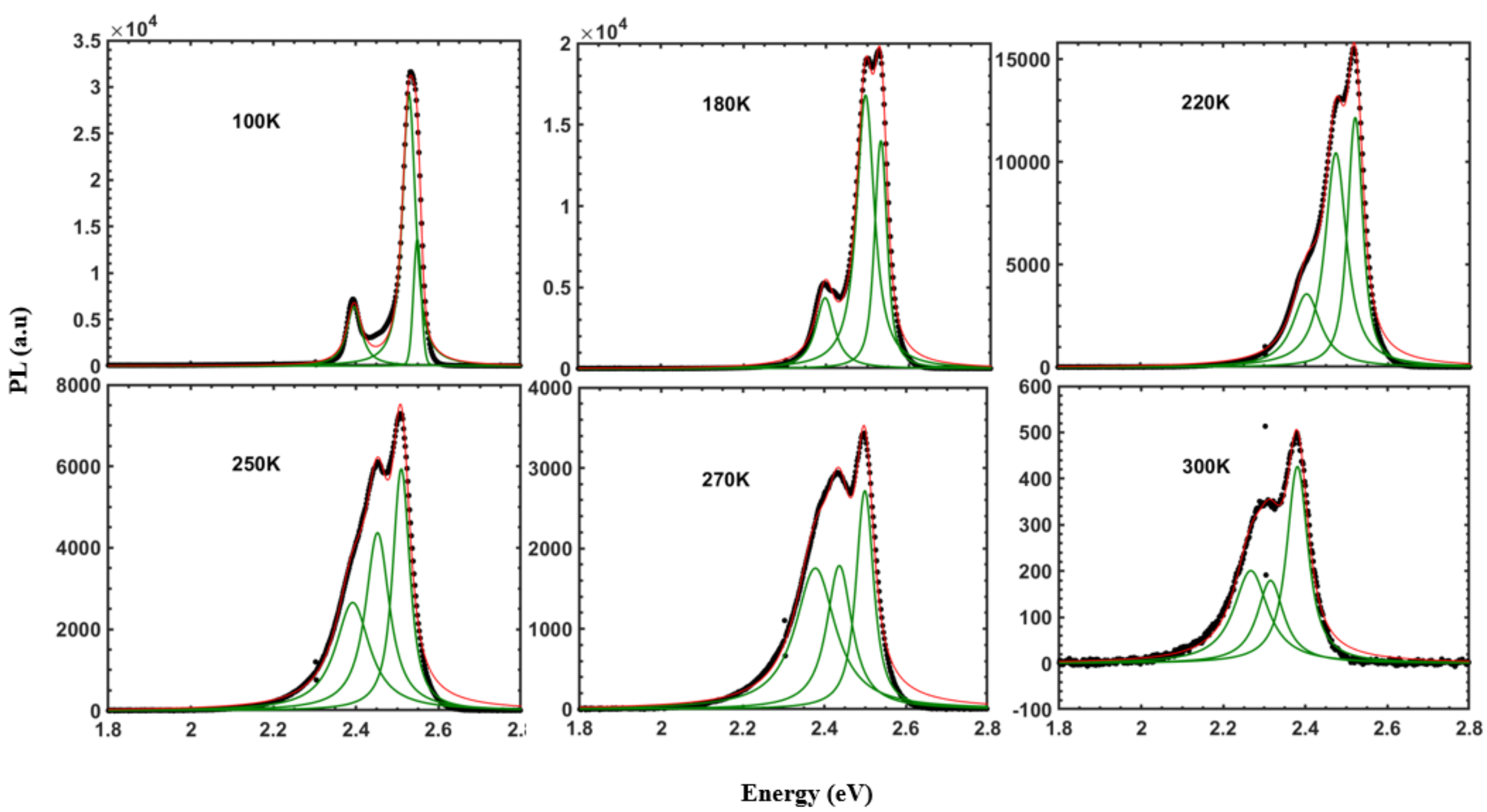

**FIG. S4** "Color" PL spectral fitted curve and peak splitting for $(BA)_2PbI_4$

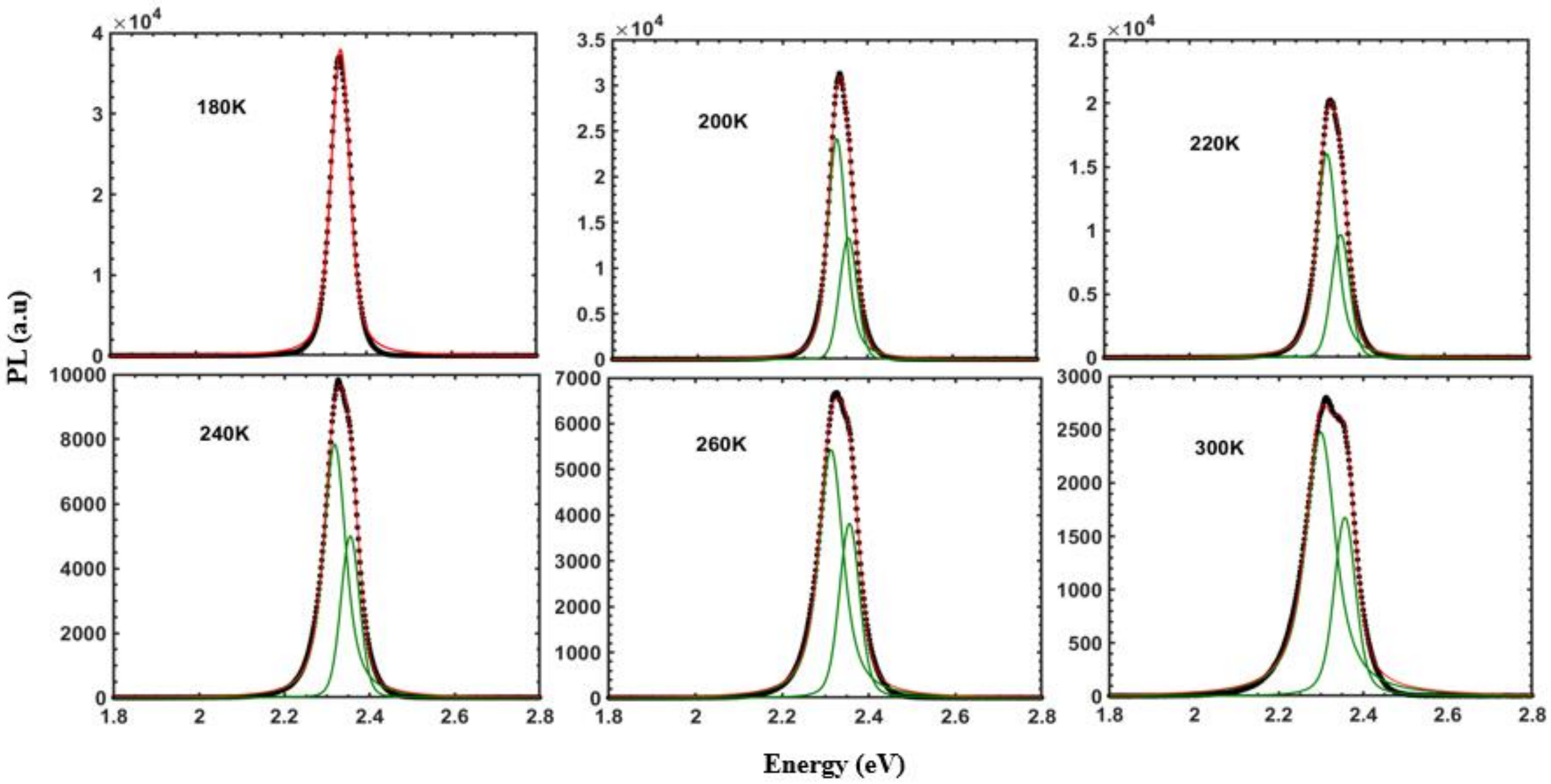

**FIG. S5** "Color" PL spectral fitted curve and peak splitting for $(PEA)_2PbI_4$